\newcommand{\useepj}{3}
\ifcase\useepj\or
\documentclass[epj,
nopacs,final
               ]{svjour}
\usepackage{amsmath,psfrag,epsfig}
\input{epjtitlepage}
\or
\documentclass[12pt]{article}
\usepackage{amsmath,psfrag,epsfig}
\def\email#1{e-mail: {\tt #1}}
\def\acknowledgement#1{\vspace{2ex}#1}
\or
\documentclass[epj,
nopacs,final
               ]{svjour}
\usepackage{amsmath,psfrag,epsfig}
\def\titlename{Regularization and supersymmetry-restoring
  counterterms in supersymmetric QCD}
\def\titlerunningname{Regularization and supersymmetry-restoring
  counterterms in SQCD}
\date{\today}
\def\authornames{
W. Hollik\thanks{\email{hollik@particle.physik.uni-karlsruhe.de}} \and
        D. St{\"o}ckinger\thanks{\email{ds@particle.physik.uni-karlsruhe.de}}}
\def\authorrunningnames{W. Hollik, D. St{\"o}ckinger}
\def\emails{hollik@particle.physik.uni-karlsruhe.de,
  ds@particle.physik.uni-karlsruhe.de} 
\def\institutename{Institut f{\"u}r Theoretische Physik, 
              Universit{\"a}t Karlsruhe,
              D--76128 Karlsruhe, Germany
              }

\fi

\renewcommand{\L}{{\cal L}}

\newcommand{\epsilonbar}{{\overline\epsilon}}

\newcommand{\chibar}{{\overline\chi}}

\newcommand{\intx}{\int d^4x}
\newcommand{\glui}{\tilde{g}}
\newcommand{\gluibar}{\overline{\tilde{g}}}
\newcommand{\cbar}{\bar{c}}
\newcommand{\dreg}{\theta_{\rm DReg}}
\newcommand{\sq}{\tilde{q}}
\newcommand{\qbar}{\bar{q}}
\newcommand{\ygluon}{{Y_G}}
\newcommand{\yglui}{{\tilde{y}_{\glui}}{}}
\newcommand{\ygluibar}{{\overline{\tilde{y}_{\glui}}}{}}
\newcommand{\yc}{{Y_c}}
\newcommand{\yq}{y}
\newcommand{\yqbar}{\bar{y}}
\newcommand{\ysq}{\tilde{y}}
\newcommand{\TF}{T(F)}
\newcommand{\CA}{C(A)}
\newcommand{\CF}{C(F)}
\newcommand{\allpsfrag}{\psfrag{y1}{$\ysq^\dagger$}
                        \psfrag{e1}{$\epsilonbar$}
                        \psfrag{e}{$\epsilon$}
                        \psfrag{s1}{$\sq^\dagger$}
                        \psfrag{y2}{$\hspace{-1ex}\yglui$}
                        \psfrag{yd}{$\yqbar$}
                        \psfrag{qd}{$\qbar$}
                        \psfrag{yg}{$\ygluon$} 
                        \psfrag{yc}{$\yc$}
                        \psfrag{y3}{$\ysq$}
                        \psfrag{g}{$\glui$}
                        \psfrag{s}{$\sq$}
                        \psfrag{c1}{$\cbar$}
                        \psfrag{c}{$c$}
                        \psfrag{q}{$q$}
                        \psfrag{G}{$G$}
                        \psfrag{(k)}{$(k)$}
                        \psfrag{(-k)}{$(-k)$}
                        \psfrag{(p)}{$(p)$}
                        \psfrag{(p1)}{$(p')$}
                       }
\def\dg#1{\frac{\delta\Gamma}{\delta#1}}

\def\dfrac#1#2{\frac{\delta{#1}}{\delta{#2}}}

\def\pslash#1{{\setbox0=\hbox{$#1$}
  \rlap{\ifdim\wd0>.7em\kern.22\wd0\else\kern.1\wd0\fi /}#1}}
\def\psl{\pslash p}
\def\ksl{\pslash k}
\def\qsl{\pslash q}

\def\GG#1{{\Gamma_{#1}}}
\allowdisplaybreaks[1]
\begin{document}

\ifcase\useepj\or
\maketitle
\or
\renewcommand{\and}{and\ }
\renewcommand{\thanks}{\footnote}

\begin{titlepage}
\begin{flushright}
KA--TP--08--2001\\
{\tt hep-ph/0103009}\\
\end{flushright}
\vspace{3ex}
\begin{center}
{\Large\bf \titlename\\}
\vspace{3ex}
{\large{\authornames}\\
  \parbox{10cm}{\small\center\em\institutename}
  }
\end{center}
\vspace{2ex}
\begin{abstract}
We calculate symmetry-restoring counterterms in supersymmetric QCD at
the one-loop level. First we determine loop corrections to the
supersymmetry and gauge transformations and find counter\-terms in such
a way that the symmetry algebra holds at the one-loop level. Then these
results are used to derive the symmetry-restoring counterterms to all
trilinear interactions. In order to obtain unique results it is
crucial to use the Slavnov-Taylor identity, which does not only
contain supersymmetric and gauge Ward identities but also describes
the symmetry algebra. In dimensional regularization this procedure
yields unique non-zero values for the counterterms. In contrast, in
dimensional reduction we find that no non-symmetric counterterms are
needed, neither for the symmetry transformations nor for the physical
interactions. For the considered cases this result constitutes a
definite test of the supersymmetry and gauge invariance of the scheme.
\end{abstract}
\end{titlepage}
\or
\thispagestyle{empty}
\twocolumn[{
\begin{flushright}
KA--TP--08--2001\\
{\tt hep-ph/0103009}\\[8ex]
\end{flushright}
\begin{flushleft}
{\noindent\LARGE\bfseries\sffamily{\titlename}\\[3ex]}
\end{flushleft}
\renewcommand{\and}{and\ }
\renewcommand{\thanks}[1]{}
{\normalsize{\authornames\\[3ex]}}
{\normalsize{\institutename\\[3ex]}}
{\normalsize{e-mail: {\tt\emails}}}
\center{\vspace{3ex}\parbox{16cm}{
\today\\[3ex]
{{\noindent
    {\bf\sffamily Abstract.\ }
We calculate symmetry-restoring counterterms in supersymmetric QCD at
the one-loop level. First we determine loop corrections to the
supersymmetry and gauge transformations and find counter\-terms in such
a way that the symmetry algebra holds at the one-loop level. Then these
results are used to derive the symmetry-restoring counterterms to all
trilinear interactions. In order to obtain unique results it is
crucial to use the Slavnov-Taylor identity, which does not only
contain supersymmetric and gauge Ward identities but also describes
the symmetry algebra. In dimensional regularization this procedure
yields unique non-zero values for the counterterms. In contrast, in
dimensional reduction we find that no non-symmetric counterterms are
needed, neither for the symmetry transformations nor for the physical
interactions. For the considered cases this result constitutes a
definite test of the supersymmetry and gauge invariance of the scheme.}}}\\[6ex]}
}]
\fi

\section{Introduction}

It has been a longstanding problem that dimensional regularization
(DReg) breaks supersymmetry. In general, this breaking necessitates
the calculation of compensating, supersymmetry-restoring
counterterms. An efficient solution would be provided by a manifestly
supersymmetric and gauge invariant regularization, but no such
regularization is known. A practically useful scheme is dimensional
reduction (DRed), which is however mathematically inconsistent and
thus cannot work at all orders \cite{Siegel79,Siegel80}. Still it has
been shown that several supersymmetric Ward identities are satisfied
in DRed \cite{CJN80} and that, as long as the
inconsistencies of DRed do not play a role, DRed is related to DReg
by a coupling constant redefinition and thus leads to equivalent
results \cite{JJR}.

At the same time there were severe difficulties to find the correct
way to renormalize supersymmetric gauge theories in a
regularization-independent way. In the Wess-Zumino gauge, which is the
gauge almost exclusively used in practical calculations, the usual way
of treating global symmetries by Ward identities was shown to fail for
supersymmetry (for an account see \cite{BrMa85}).

By now these difficulties have been solved
for exact supersymmetry \cite{White92a,MPW96a,HKS99} as
well as for the case of softly broken supersymmetry
\cite{MPW96b,SYM}. In particular, a consistent set of symmetry
identities (Ward- and Slavnov-Taylor identities) has 
been found that provides an unambiguous definition of the theories.

There is a major difference between the supersymmetric Slavnov-Taylor
identity found in \cite{White92a} and the Ward identities considered
in \cite{CJN80,BrMa85}. For Green functions, the Ward identities can
be written as 
\begin{eqnarray}
\delta_{\rm susy} \langle T\;\phi_1\ldots\phi_n\rangle
& = & \langle T\Delta\; \phi_1\ldots\phi_n\rangle \ ,
\label{SusyWI}
\end{eqnarray}
where $\delta_{\rm susy}$ denotes the infinitesimal supersymmetry
transformations of the fields and the composite operator $\Delta$ is
due to the supersymmetry 
breaking of the gauge fixing in the Wess-Zumino gauge.
The symmetry transformations are generally non-linear composite
operators receiving quantum corrections that have to be renormalized,
and the meaning of (\ref{SusyWI}) is the invariance of the theory
under these renormalized transformations. But the invariance expressed
by (\ref{SusyWI}) does not necessarily correspond to supersymmetry. It
only does if the renormalized transformations satisfy the 
supersymmetry algebra, and only the Slavnov-Taylor identity contains
both the invariance of the theory and the algebra. Indeed,
the Slavnov-Taylor identity  determines the structure constants of the
symmetry algebra and thus governs the renormalization of the symmetry
operators. As a consequence, only the Slavnov-Taylor identity yields
unique results for the non-supersymmetric counterterms or provides
definite tests of the supersymmetry of a calculation. 

The purpose of this article is twofold. First we develop the
Feynman rules involving all the ghost fields and external fields that
appear in the Slavnov-Taylor identity. Then we use the Slavnov-Taylor
identity to determine supersymmetry-restoring counterterms in
dimensional regularization and to check whether a given regularization
such as DRed
preserves supersymmetry at one-loop order. We do this in
supersymmetric QCD with soft breaking, a model of particular
phenomenological interest with generally large quantum
corrections. The Slavnov-Taylor identity not only describes
supersymmetry but also gauge invariance. Taking into account all its
consequences for the symmetries and the symmetry algebra, we calculate
all symmetry-restoring counterterms to the trilinear interactions and
the required counterterms to the symmetry transformations. 

Section \ref{Sec:STI} sets the basis for our calculations. Since the
Slavnov-Taylor identity is the key relation, it is briefly reviewed,
with emphasis on the meaning of the involved ghost fields and of the
special cases we need in later course. 

In sec.\ \ref{Sec:Calculation} we derive and discuss the counterterms
to all trilinear interactions and the necessary counterterms to the
symmetry transformations; in sec.\ \ref{Sec:Conclusions} we present
our conclusions. In App.\ \ref{Sec:FR} the Feynman rules are listed,
in particular the ones involving the ghost fields. We need them to
calculate the loop corrections to the symmetry transformation operators. 
The rest of the appendices is devoted to the explicit form of the
Lagrangian, the BRS transformations, and the one-loop results of the
necessary vertex functions.

\section{The Slavnov-Taylor identity of supersymmetric QCD}
\label{Sec:STI}

In this section we briefly review the Slavnov-Taylor identity of
supersymmetric QCD \cite{White92a,MPW96a,HKS99,MPW96b,SYM}, since it
is the mathematical expression for the symmetries of the model. Its
knowledge is necessary for testing whether a regularization preserves
the symmetries as well as for the determination of
supersymmetry-restoring counterterms.

\subsection{BRS invariance and gauge fixing}


The usual gauge fixing term of supersymmetric QCD
\begin{eqnarray}
\L_{\rm fix} & = & -\frac{1}{2\xi} (\partial_\mu G^\mu_a)^2
\label{GaugeFix}
\end{eqnarray}
does not only break gauge invariance but also supersymmetry because it
contains only the gluon $G^\mu_a$ but not the gluino. As a
consequence it is very useful to treat gauge invariance and
supersymmetry simultaneously using combined BRS
transformations. Then not only the Faddeev-Popov ghosts are needed but
also supersymmetry ghosts.

Actually, three kinds of ghost fields are introduced: the
Faddeev-Popov ghosts $c_a(x)$ that correspond to gauge transformations
and are fermionic scalar fields, supersymmetry ghosts $\epsilon$
(bosonic Majorana spinors), and translational ghosts $\omega^\mu$
(fermionic vectors). Translations cannot be treated separately since
they arise as anticommutators of supersymmetry transformations. Among
these ghost fields only the $c_a(x)$ are dynamical fields, whereas
$\epsilon$ and $\omega^\mu$ are space-time independent constants since
the corresponding symmetry transformations are global.

The explicit form of the BRS operator $s$ is listed in the
appendix. As a crucial property of the BRS transformations, the
transformations of the ghosts are determined by the structure 
constants of the symmetry algebra. Hence, knowing these BRS
transformations is equivalent to knowing all the (anti-)commutators of
the symmetry generators. It is this property that renders the BRS
operator nilpotent, so that $s^2=0$ up to equations of motions. 

Using the nilpotency of the BRS operator it is possible to write down
a BRS-invariant gauge-fixing and ghost term: For the usual $\xi$-gauge
one has
\begin{eqnarray}
\Gamma_{\rm fix,\ gh} & = & \intx\L_{\rm fix,\ gh}\ ,
\nonumber\\
\L_{\rm fix,\ gh} & = & 
 s[\bar{c}_a (f_a + \frac{\xi}{2}B_a)]
\nonumber\\
& = &
B_a f_a + \frac{\xi}{2} B_a^2 
- \bar{c}_a\partial_\mu (D^\mu c)_a
\nonumber\\&&{}
- \bar{c}_a\partial^\mu(\epsilonbar\gamma_\mu\glui_a) 
+ \frac12\xi i \epsilonbar\gamma^\nu\epsilon
  (\partial_\nu\bar{c}_a)\bar{c}_a 
\ ,
\label{GaugeFixing}
\end{eqnarray}
with $f_a = \partial_\mu G^\mu_a$ and the Faddeev-Popov antighosts
$\cbar_a(x)$. For diagrammatic calculations it is customary to
eliminate the auxiliary fields $B_a$, yielding the usual gauge fixing
term (\ref{GaugeFix}). Note that the supersymmetry breaking of this
gauge fixing term necessitates compensating terms involving the
$\epsilon$ ghosts. 

\subsection{Slavnov-Taylor identity}

At the tree level, BRS invariance can be expressed in the following
way: 
\begin{eqnarray}
0 & = & \intx\ s\varphi_i \dfrac{\Gamma_{\rm inv}}{\varphi_i}\ ,
\label{BRSInvariance}
\end{eqnarray}
where $\Gamma_{\rm inv}$ is a BRS invariant action and the sum runs
over all fields of the model. At the quantum level, the BRS
transformations have to be treated as composite operators. The
non-linear composite operators receive loop corrections and have to be
renormalized in the same way as the Green functions. It is most
convenient to perform the renormalization of these composite symmetry
operators together with the renormalization of the Lagrangian. To do
that we couple the non-linear BRS transformations $s\varphi_i$ to
external sources $Y_i$ according to the scheme\footnote{The 
  minus sign applies for complex conjugate fields. The
  reason is the reality of the BRS operator
  leading to the rule $(s Bos)^\dagger=s(Bos)^\dagger$,
  $(sFer)^\dagger = - s(Fer)^\dagger$.}
\begin{eqnarray}
s\varphi_i &\longrightarrow & \pm\dfrac{\Gamma_{\rm cl}}{Y_i} + {\cal
 O}(Y_i)\ ,\\
\langle s\varphi_i\rangle_J & \longrightarrow & \pm\dfrac{\Gamma}{Y_i}
 + {\cal O}(Y_i)
\end{eqnarray}
at the classical and at the quantum level. The resulting classical
action $\Gamma_{\rm cl}$ is written down in App.\ \ref{Sec:L}. 
$\Gamma$ denotes the renormalized effective action, the generating
functional of one-particle irreducible vertex 
functions.\footnote{The sources $J_i=-\dg{\varphi_i}$  for the fields
  are to be understood as the usual sources in the functional
  integral.} Clearly, the expectation values of products of operators
differ from the products of the individual expectation values. This
reflects the appearance of non-trivial loop corrections to the
non-linear composite operators. 

Now it is possible to write down the Slavnov-Taylor identity. It
expresses the invariance under the loop-corrected BRS transformations
(which incorporate gauge and supersymmetry transformations and
translations) and the fact that the loop-corrected transformations
still satisfy the desired symmetry algebra. The Slavnov-Taylor
identity reads:
\begin{eqnarray}
S(\Gamma) & = & 0\ ,\nonumber\\
S(\Gamma) & = & S_0(\Gamma)+S_{\rm soft}(\Gamma)\ ,
\end{eqnarray}
with  the part corresponding to unbroken supersymmetry,
\begin{eqnarray}
S_0(\Gamma) & = & \int
\Bigl(\dg{\ygluon_{a\mu}}\dg{G^\mu_a} + \dg{\glui_a}\dg{\ygluibar_a}
\nonumber\\&&{}
+\sum_{k=L,R}\Bigl(\dg{\ysq_k}\dg{\sq_k}
        -\dg{\ysq^\dagger_k}\dg{\sq^\dagger_k}\Bigr)
\nonumber\\&&{}
+\dg{q}\dg{\yqbar}-\dg{\yq}\dg{\qbar}
\nonumber\\&&{}
+\dg{\yc_a}\dg{c_a} + s\bar{c}_a\dg{\bar{c}_a}
+sB_a\dg{B_a}\Bigr)
\nonumber\\&&{}
 + s\omega^\mu\dg{\omega^\mu}\ ,
\end{eqnarray}
and the part describing the soft breaking,
\begin{eqnarray}
S_{\rm soft} & = & \int\Bigl(
sa\dg{a}+sa^\dagger\dg{a^\dagger}
+s\chi\dg{\chi}
\nonumber\\&&{}
+sf\dg{f}+sf^\dagger\dg{f^\dagger}\Bigr)\ .
\end{eqnarray}
In the Slavnov-Taylor operator all fields of the model appear: The
gluons $G^\mu_a$ and the gluinos $\glui_a$, the quark $q$ and the squarks
$\sq_{L,R}$ as well as the ghost fields, and the corresponding sources.
Due to the squark mixing, the mass eigenstates $\sq_{1,2}$ are in general
different from the interaction eigenstates $\sq_{L,R}$. We write the
relation as follows,
\begin{eqnarray}
\sq_k & = & S_{kL} \sq_L + S_{kR}\sq_R\ ,\nonumber\\
\ysq_k & = & S^*_{kL} \ysq_L + S^*_{kR} \ysq_R
\end{eqnarray}
with a unitary matrix $S$ diagonalizing the tree level squark mass
matrix. 
$S_{\rm soft}(\Gamma)$ involves the auxiliary chiral supermultiplet
$(a,P_L\chi,\hat{f}=f+f_0)$ and its hermitian conjugate. The main
property of this auxiliary multiplet is the constant piece $f_0$ that
acts like a vacuum expectation value of $\hat{f}$ and generates the
soft-breaking terms while permitting a fully supersymmetric
formulation of the model.

\subsection{Important special cases}

For our later applications several special cases of the Slavnov-Taylor
identity are particularly important. First it can be used to describe
gauge invariance by taking the derivative with respect to the
Faddeev-Popov ghost $c_a$ and setting all ghost fields and sources to
zero (``gh=0''):
\begin{eqnarray}
0 & = & \dfrac{S(\Gamma)}{c_a}
\nonumber\\
\Rightarrow\quad 0 & = & \dg{c_a\delta Y_i}\dg{\varphi_i} +
s\cbar_b\dg{c_a\delta \cbar_b} \Big|_{\rm gh=0}\
.
\end{eqnarray}
The functions $\delta\Gamma/\delta c_a\delta Y_i$ are the
loop-corrected gauge transformations of $\varphi_i$, and the last term
in this identity is due to gauge fixing. 
Similarly, the Slavnov-Taylor identity can be used to describe
supersymmetry by taking the derivative with respect to the
supersymmetry ghost and setting all ghost fields and sources to
zero:
\begin{eqnarray}
0 & = & \dfrac{S(\Gamma)}{\epsilon}
\nonumber\\
\Rightarrow\quad 0 & = & \dg{\epsilon\delta Y_i}\dg{\varphi_i} +
\dfrac{s\chi}{\epsilon}\dg{\chi} \Big|_{\rm gh=0}\ .
\end{eqnarray}
Here the first term is the supersymmetry transformation of
$\varphi_i$, the last term is due to the soft supersymmetry
breaking. When the auxiliary fields $a,\chi,f$ are set to zero,
one has $\delta s\chi/\delta\epsilon=\sqrt2(P_L-P_R)f_0$ with the
constant $f_0$. Apart from the soft-breaking term, these identities
are similar to the supersymmetric Ward identities (\ref{SusyWI}),
rewritten for one-particle irreducible Green functions.

As already noted, the Slavnov-Taylor identity also describes the
symmetry algebra. This information is very important in order to
guarantee that the loop-corrected symmetry transformations
still satisfy the (anti-)commutation relations that define
supersymmetry and SU(3)-gauge invariance. We can extract information
about the algebra by taking derivatives of the following kind and
setting then all ghost fields and sources to zero:
\begin{eqnarray}
0 & = & \dfrac{^3 S(\Gamma)}{\epsilon\delta\epsilonbar\delta Y_j}
\nonumber\\
\Rightarrow\quad 0 & = &
\dg{\epsilonbar\delta Y_i}\dg{\epsilon\delta Y_j\delta\varphi_i} +
\dg{\epsilon\delta Y_i}\dg{\epsilonbar\delta Y_j\delta\varphi_i}
\nonumber\\&&{}
+ \dg{\epsilon\delta\epsilonbar\delta Y_j\delta Y_i}\dg{\varphi_i} 
+ \dg{\epsilon\delta\epsilonbar\delta\yc_a}\dg{Y_j\delta c_a}
\nonumber\\&&{}
 +
\dfrac{^2 s \omega^\mu}{\epsilon\delta\epsilonbar}\dg{Y_j\delta
  \omega^\mu} \Big|_{\rm gh=0}\ . 
\end{eqnarray}
The first two terms express the anticommutator of the supersymmetry
transformations of $\varphi_j$ into $\varphi_i$ and of $\varphi_i$;
the remaining terms express the right-hand side of the supersymmetry
algebra 
\begin{eqnarray}
\{Q,\bar{Q} \} & = &\mbox{equations of motion}+\mbox{gauge transformations}
\nonumber\\&&{}+ 2\gamma^\mu P_\mu 
\ ,
\end{eqnarray}
where the coefficient of the translational part is fixed by
\begin{eqnarray}
\dfrac{^2 s \omega^\mu}{\epsilon\delta\epsilonbar}= 2\gamma^\mu\ .
\end{eqnarray}

Later we will use these identities taking further derivatives and
setting all fields to zero.

\subsection{Definition of the model}
\label{Sec:Definition}

The Slavnov-Taylor identity is not the only symmetry identity in
supersymmetric QCD. For an unambiguous definition of supersymmetric
QCD we need four symmetry identities. We require them to be satisfied
by the renormalized effective action $\Gamma$ (see \cite{SYM}):
\begin{itemize}
\item The Slavnov-Taylor identity $S(\Gamma)=0$ expressing gauge
  invariance, supersymmetry and translational invariance.
\item The gauge fixing condition $\dg{B_a}=\frac{\delta\Gamma_{\rm
      fix}}{\delta B_a}=f_a+\xi B_a$ expressing the
  non-renormalization of the gauge fixing terms.
\item The translational ghost equation $\dg{\omega^\mu} = \frac{\delta\Gamma_{\rm
      ext}}{\delta \omega^\mu}$ ($\Gamma_{\rm ext}$ is defined in
  App.\ \ref{Sec:Construction}) meaning that the terms involving
  $\omega^\mu$ do not receive quantum corrections.
\item Global $SU(3)$ invariance and invariance under continuous $R$
  transformations with the $R$-weights defined in tab.\
  \ref{Tab:QuantumNumbers} and $CP$ invariance. 
\end{itemize} 
\ifcase\useepj\or\def\whichtable{table}\or\def\whichtable{table*}\or\def\whichtable{table}\fi
\begin{\whichtable}[tbh]
\begin{displaymath}
\begin{array}{|c||c|c|c|c|c|c|c|c|c|c|c|c|c|c|c|c|c|c|}
\hline
\varphi & G_a^\mu  & P_L\glui_a &  
\sq_{L},\sq_R^\dagger &
q & a & P_L\chi & \hat{f} &  
 c_a & P_L\epsilon & 
\omega^\nu & \bar{c}_a & B_a 
\\ \hline
R   & 0  & 1 & 1  &  0  & 0 & -1 & -2 & 0 & 1  & 0 & 0 & 0 \\ \hline 
Q_c & 0  & 0 &  0 &  0  & 0 & 0  &  0 &+1 & +1 &+1 &-1 & 0 \\ \hline
GP  & 0  & 1 &  0 &  1  & 0 & 1  &  0 & 1 & 0  & 1 & 1 & 0 \\ \hline
dim & 1  &3/2&  1 & 3/2 & 0 & 1/2&  1 & 0 &-1/2&-1 & 2 & 2 \\ \hline
\end{array}
\end{displaymath}
\caption{Quantum numbers. $R,Q_c,GP,dim$ denote $R$-weight and ghost
  charge, Grassmann parity and the mass dimension, respectively. The
  $R$-weights of the right-handed parts $P_R\glui$, $P_R\epsilon$, $P_R\chi$ of
  the Majorana spinors are opposite to the ones of the left-handed
  parts. The quantum numbers of the external fields $Y_i$  can be
  obtained from the requirement that the products $Y_i s\varphi_i$ are
  neutral, bosonic and have $dim=4$. The commutation rule for two
  general fields is $\varphi_1\varphi_2 = (-1)^{GP_1 GP_2} \varphi_2\varphi_1$.}
\label{Tab:QuantumNumbers} 
\end{\whichtable}

\section{Determination of symmetry-restoring counterterms}
\label{Sec:Calculation}

In general, regularization schemes break the defining symmetry
identities. Since supersymmetric QCD is anomaly free it is always
possible to restore the symmetries by adding appropriate
counterterms $\Gamma_{\rm non-sym}$ that break the symmetries by
themselves. In this section we determine such counterterms at one-loop
order.

A key issue in this
determination is the uniqueness of the counterterms. It is not
sufficient to calculate a counterterm by considering only one symmetry
identity because all symmetry identities have to be satisfied
simultaneously. If, however, a counterterm is determined uniquely by a
certain set of symmetry identities, then it is the simultaneous
solution to all identities. 

The strategy in this section is the following:
\begin{itemize}
\item Calculate the counterterms to the supersymmetry transformations
  of the gluon and the gluino using identities expressing the
  supersymmetry and the supersymmetry algebra.
\item Determine the counterterms to the supersymmetry transformations
  of the squarks and the quark in the same way.
\item Derive the counterterm to the $\sq\glui q$ interaction using a
  supersymmetry identity together with the counterterms calculated
  before. 
\item Determine the counterterms to all three-particle gauge
  interactions and the relevant gauge transformations using identities
  expressing gauge invariance and the SU(3) algebra.
\item Cross-check the result for the gauge interactions using a
  supersymmetry identity relating the $G_\rho G_\nu G_\mu$ and the
  $\glui\glui G_\mu$ interactions.
\end{itemize}

\subsection{Parametrization of the counterterms}
\label{Sec:Parametrization}

Counterterms can be divided into symmetric and non-symmetric ones,
\begin{eqnarray}
\Gamma_{\rm ct} & = & \Gamma_{\rm sym} + \Gamma_{\rm non-sym}\ .
\end{eqnarray}
The symmetric counterterms $\Gamma_{\rm sym}$ do not destroy any
symmetry identity. They can be obtained from the classical action 
\begin{eqnarray}
\Gamma_{\rm cl} \longrightarrow
\Gamma_{\rm cl} + \Gamma_{\rm sym}
\end{eqnarray}
by infinitesimal renormalization transformations for fields and
parameters \cite{MPW96b,SYM}: 
\begin{align}
G^\mu& \to \sqrt{Z_G} G^\mu,&
B& \to \sqrt{Z_G}^{-1} B,
\nonumber\\
\cbar & \to \sqrt{Z_G}^{-1}\cbar,  &
\xi & \to  Z_G\xi 
,\nonumber\\
\glui & \to  \sqrt{Z_{\glui}} \glui,&
c & \to  \sqrt{Z_c}c,\nonumber\\
P_{L,R}q & \to  \sqrt{Z_{q_{L,R}}}P_{L,R}q,&
\sq_{L,R} & \to  \sqrt{Z_{L,R}}\sq_{L,R},\nonumber\\
Y_i & \to  \sqrt{Z_i}^{-1} Y_i,\nonumber\\
g & \to  g+\delta g,&
m_i & \to  m_i+\delta m_i ,
\end{align}
where $m_i$ denotes all mass parameters of the theory including the
soft parameters.\footnote{It is
  also possible to 
  perform a matrix valued renormalization transformation of the
  squarks. This is important if complete on-shell renormalization
  conditions are desired, but for our concern the difference is not
  relevant.} Owing to this structure the contributions of
$\Gamma_{\rm sym}$ to all self-energies and to the interaction vertex
$\sq\sq G_\mu$ have completely arbitrary coefficients that can be
chosen at will. All other contributions are functions of this
choice. Formally this feature can be expressed as
\begin{eqnarray}
\Gamma_{\rm sym} & = & \sum_i \delta_{\rm sym}{}_i^{(1)} {\cal O}_i^{(1)}
                     + \sum_i \delta_{\rm sym}{}_i^{(2)} {\cal O}_i^{(2)}\ ,
\end{eqnarray}
where the operators ${\cal O}^{(1)}_i$ correspond to the self-energies
and the $\sq\sq G_\mu$ vertex and the coefficients $\delta_{\rm
  sym}{}_i^{(2)}$ are functions of the $\delta_{\rm sym}{}_i^{(1)}$.

The second contribution in $\Gamma_{\rm ct}$ are the non-symmetric but
symmetry-restoring counterterms $\Gamma_{\rm non-sym}$. Since the contribution
of $\Gamma_{\rm   sym}$ to the coefficients of the operators ${\cal
  O}_i^{(1)}$ is already completely arbitrary, it is possible to
assume without loss of generality that $\Gamma_{\rm non-sym}$ has the form
\begin{eqnarray}
\Gamma_{\rm non-sym} & = & \sum_i\delta^{(2)}_i
{\cal O}_i^{(2)}\ ,
\end{eqnarray}
which means that $\Gamma_{\rm non-sym}$ does not contain contributions
to the self-energies and the $\sq\sq G_\mu$ vertex.

This parametrization of the counterterms we will use in the following
calculations. Since it is completely general, our results are valid
independently of the symmetric counterterms. Hence they hold for all
renormalization schemes such as the $\overline{MS}$ or the on-shell
scheme. But since this parametrization avoids redundancies it is
particularly well suited  for a transparent discussion. By
construction, the symmetric counterterms drop from all symmetry
identities, and therefore the number of unknown counterterms in the
identities is minimized.

There is only one restriction we impose on $\Gamma_{\rm
  non-sym}$. Since all common regularization schemes preserve
global SU(3)- and $CP$-invariance, we do not admit counterterms
in $\Gamma_{\rm non-sym}$ that break these symmetries.

\subsection{Gluon and gluino self energies and supersymmetry
  transformations}
\label{Sec:GgSelfEnergies}

Taking the derivative of the Slavnov-Taylor identity
\begin{eqnarray}
0 & = & \frac{\delta^3 S(\Gamma)}
             {\delta G_{b\mu} \delta \epsilon \delta \gluibar_a}
\end{eqnarray}
and setting all fields to zero
we obtain an identity relating the gluon and gluino self
energies,\footnote{In the rest of this section $\Gamma$ denotes the
  one-loop effective action including the contributions of
  $\Gamma_{\rm non-sym}$. The symmetric counterterms do not appear
  since they drop from every symmetry identity.}
\begin{eqnarray}
0 & = & \GG{\epsilon\gluibar_a \ygluon^{\nu}_c}(-q,q)
        \GG{G_{b\mu} G_{c\nu}}(q,-q)
\nonumber\\&&{}
      - \GG{\glui_c\gluibar_a}(q,-q)
        \GG{G_{b\mu}\epsilon\ygluibar_c}(q,-q)
\nonumber\\&&{}
      + \frac{\delta s\chi}{\delta\epsilon}
        \GG{G_{b\mu}\gluibar_a\chi}(q,-q)
\ .
\label{GgSTIOrig}
\end{eqnarray}
In this identity the Green functions corresponding to the
loop-corrected supersymmetry transformations of the gluon and the
gluino are involved, and the identity determines the ratio of these
two supersymmetry transformations. The last term is due to the soft
supersymmetry breaking. 

The notation $\GG{\varphi_1\ldots\varphi_n}$ means the
one-particle irreducible vertex function with external
$\varphi_1$\ldots$\varphi_n$ fields
\begin{eqnarray}
\GG{\varphi_1\ldots\varphi_n} & = & \dfrac{^n
  \Gamma}{\varphi_1\ldots\delta\varphi_n}\Big|_{\varphi_i=0}\ ,
\end{eqnarray}
and the momentum arguments denote the incoming momenta (note that
$\epsilon$ is a constant and thus does not carry a momentum).

In order to obtain the counterterms it is sufficient to consider the
high-momentum limit. Then the soft-breaking term is negligible, and
the results from App.~\ref{Sec:Oneloop} yield 
(the quantities $\CA$ and $\TF$ are defined in App.\ \ref{Sec:SU3},
and the one-loop functions $B_0$ are defined in App.\ \ref{App1LInt})
\begin{eqnarray}
0 & = & (-\gamma^\mu q^2+\qsl q^\mu)\times
\nonumber\\&&{}
\biggl( \Bigl[1 +\delta_{\ygluon\glui\epsilon}+
        \frac{\alpha_s \CA}{4\pi}\Bigl(-B_0-\frac13\dreg + 
        B_0
        \Bigr) 
\nonumber\\&&{}
        + \frac{\alpha_s \TF}{4\pi}2B_0\Bigr]
\nonumber\\&&{}
      - \Bigl[1 + \delta_{G\epsilon\yglui}+\frac{\alpha_s \CA}{4\pi}\Bigl(B_0 - 1\dreg 
          - B_0\Bigr)
\nonumber\\&&{}
            + \frac{\alpha_s \TF}{4\pi}2B_0 
        \Bigr]
\biggr)
\nonumber\\
& = & (-\gamma^\mu q^2+\qsl q^\mu) \times
\nonumber\\&&{}
\Bigl( \delta_{\ygluon\glui\epsilon}-\delta_{G\epsilon\yglui}
        +\frac{\alpha_s \CA}{4\pi} \frac23\dreg  \Bigr)
\ .
\end{eqnarray}
Thus, in dimensional regularization ($\dreg=1$) this identity is not
satisfied on the regularized level; one has to choose non-vanishing
values for the counterterms
\begin{eqnarray}
\delta_{\ygluon\glui\epsilon} 
                           - \delta_{G\epsilon\yglui}
& = & -\frac23\frac{\alpha_s \CA}{4\pi}\dreg 
\label{GgSTI}
\end{eqnarray}
to the supersymmetry transformations of the gluon and the gluino.

In order to determine the individual counterterms we derive an
identity corresponding to the supersymmetry algebra:
\begin{eqnarray}
0 & = & \frac{\delta^4 S(\Gamma)}
             {\delta G^\nu_b \delta\epsilon\delta\epsilonbar
              \delta \ygluon^{\mu}_a}
\nonumber\\
\Rightarrow\quad 0 & = & 
\GG{\epsilon\epsilonbar \ygluon^{\mu}_a \ygluon^{\rho}_c}
\GG{G^\nu_b G_{c\rho}}
\nonumber\\&&{}
+ \GG{\ygluon^{\mu}_a\glui_c\epsilonbar}
\GG{G^\nu_b\epsilon\ygluibar_c}
- \GG{G^\nu_b\yglui_c\epsilonbar}
\GG{\ygluon^{\mu}_a\epsilon\gluibar_c}
\nonumber\\&&{}
+ \GG{G_b^\nu\epsilon\epsilonbar \yc_c}
\GG{\ygluon^{\mu}_a c_c}
\nonumber\\&&{}
+ \frac{\delta^2 s\omega^\rho}{\delta\epsilon\delta\epsilonbar}
\GG{G^\nu_b \ygluon^{\mu}_a \omega^\rho}
+ 2\frac{\delta s\chi}{\delta\epsilon}
\GG{G^\nu_b\epsilonbar \ygluon^{\mu}_a\chi}
\ ,
\label{GgAlgebra}
\end{eqnarray}
which yields in the high-momentum limit
\begin{eqnarray}
0 & = & 
-2(q_\mu \gamma_\nu - \qsl g_{\mu\nu})\delta_{ab}
\Bigl(1+\delta_{G\epsilon\yglui}+
 \delta_{\ygluon\glui\epsilon}\Bigr)
\nonumber\\&&{}
+2(q_\mu \gamma_\nu )\delta_{ab}
\Bigl(1+\delta_{c\ygluon}+\delta_{G\epsilon\epsilonbar\yc}\Bigr)
\nonumber\\&&{}
-2\qsl g_{\mu\nu}\delta_{ab}
\ .
\end{eqnarray}
The physical meaning of this identity is the constraint of the
supersymmetry algebra on the product of the transformations of the
gluon into the gluino and backwards --- correspondingly it determines
the sum of the two counterterms. In contrast, the previous identity
(\ref{GgSTIOrig}) determines the ratio of the two supersymmetry
transformations and therefore the difference of the counterterms.

Hence, taken together both identities
lead to a unique value for the counterterms
\begin{eqnarray}
\delta_{\ygluon\glui\epsilon} & = & - \delta_{G\epsilon\yglui}
= -\frac{\alpha_s \CA}{4\pi}\frac13\dreg\ ,
\\
\delta_{G\epsilon\epsilonbar\yc} & = & -\delta_{c\ygluon}\ .
\end{eqnarray}
The counterterms $\delta_{G\epsilon\epsilonbar\yc}$ and
$\delta_{c\ygluon}$ we will not need in the following.

This result is a simple illustration of the discussion at the
beginning of this section. Apparently there are infinitely many
different counterterms that solve (\ref{GgSTI}) or equivalently
(\ref{GgSTIOrig}), i.e.\ restore the 
gluon-gluino identity. Only one particular choice, however, also
solves the second identity (\ref{GgAlgebra}). Therefore it is
essential to take into account the identities that correspond to the
supersymmetry algebra. 

\subsection{Quark and squark self energies and supersymmetry
  transformations}

Using the following derivative of the Slavnov-Taylor identity we
obtain an identity relating the quark and squark self
energies:\footnote{We suppress the colour indices since the 
  following identities are trivial in colour space.}
\begin{eqnarray}
0 & = & \frac{\delta^3 S(\Gamma)}{\delta
  q\delta\sq^\dagger_i\delta\epsilonbar} 
\nonumber\\
\Rightarrow\quad 0 & = &
\sum_{j=1,2}\GG{q\epsilonbar\ysq_j}\GG{\sq^\dagger_i\sq_j}
-\GG{\sq_i^\dagger \yq\epsilonbar}\GG{q\qbar}
\nonumber\\&&{}
+\frac{\delta s\chibar}{\delta\epsilonbar}
\GG{q\sq^\dagger_i\chibar}\ .
\end{eqnarray}
In the high-momentum limit this reduces to
(the quantity $\CF$ is defined in App.\ \ref{Sec:SU3})
\begin{eqnarray}
0 & = & \sqrt2 q^2(S_{iL}P_L - S_{iR}P_R)
 \Bigl(1+\delta_{\ysq\epsilon q}+\frac{\alpha_s \CF}{4\pi}
 B_0\Bigr)
\nonumber\\&&{}
- \sqrt2 q^2(S_{iL}P_L - S_{iR}P_R)\times
\nonumber\\&&{}
 \Bigl(1+\delta_{\yq\sq\epsilon}+\frac{\alpha_s \CF}{4\pi}
 (2B_0-1\dreg - (B_0)) \Bigr)
\ ,
\end{eqnarray}
which is satisfied provided the counterterms fulfil
\begin{eqnarray}
\delta_{\ysq\epsilon q}-\delta_{\yq\sq\epsilon}
=
 - \frac{\alpha_s \CF}{4\pi}\dreg
\ .
\end{eqnarray}
Again, these are counterterms to the supersymmetry transformations,
and the considered identity only fixes their difference.

As in the gluon/gluino case we need an additional identity
corresponding to the supersymmetry algebra to find unique values for
these counterterms, given by
\begin{eqnarray}
0 & = & \frac{\delta^4 S(\Gamma)}
             {\delta\sq_j\delta\epsilon\delta\epsilonbar\delta\ysq_i}
\nonumber\\
\Rightarrow\quad 0 & = &
-\GG{\epsilon\epsilonbar\ysq_i\ysq^\dagger_k}
\GG{\sq_j\sq_k^\dagger}
\nonumber\\&&{}
+\GG{\ysq_i q\epsilonbar}\GG{\sq_j\epsilon \yqbar}
-\GG{\sq_j \yq^C\epsilonbar}\GG{\ysq_i \epsilon\qbar^C}
+ \frac{\delta^2 s\omega^\rho}{\delta\epsilon\delta\epsilonbar}
\GG{\sq_j\ysq_i\omega^\rho}
\nonumber\\&&{}
+ \frac{\delta s\chi}{\delta\epsilon}
\GG{\sq_j\epsilonbar\ysq_i\chi}
+ \frac{\delta s\chi}{\delta\epsilonbar}
\GG{\sq_j\epsilon\ysq_i\chi}
\ .
\end{eqnarray}
In the high-momentum limit only the second, third and fourth term
  contribute, with the result\footnote{The values 
  for the vertex functions involving $\yq^C$, $\qbar^C$ can easily be
  inferred from the corresponding ones involving $\yqbar$, $q$
  using the flipping rules of \cite{DEHK92}.}
\begin{eqnarray}
0 & = & 2\qsl \delta_{ij} \Bigl(1+\delta_{\ysq\epsilon q}
                                 +\delta_{\yq\sq\epsilon}\Bigr)
-2\qsl \delta_{ij}
\ .
\end{eqnarray}
The unique solution for the counterterms is thus
\begin{eqnarray}
\delta_{\ysq\epsilon q}=-\delta_{\yq\sq\epsilon}=-\frac{\alpha_s \CF}{4\pi}\frac12\dreg
\ .
\end{eqnarray}

\subsection{Gluino-Quark-Squark vertex}

One very important consequence of supersymmetry is the relation
between the interactions of quarks and squarks with gluons and
gluinos. This relation can be expressed by the following identity:
\begin{eqnarray}
0 & = & \frac{\delta^4 S(\Gamma)}
        {\delta\sq_L\delta\sq_L^\dagger\delta\glui_{aR}\delta\epsilonbar}
\nonumber\\
\Rightarrow\quad 0 & = &
\GG{\glui_{aR}\epsilonbar \ygluon^{\mu}_c}(k,-k)
\GG{\sq_L\sq_L^\dagger G_{c\mu}}(p,-p',k)
\nonumber\\&&{}
- \GG{\sq_L\sq_L^\dagger\yglui_c\epsilonbar}(p,-p',k)
\GG{\glui_{aR}\gluibar_c}(k,-k)
\nonumber\\&&{}
+ \sum_{j=1,2}
\Bigl(\GG{\glui_{aR}\epsilonbar\sq_L\ysq_j}(k,p,-p')
\GG{\sq^\dagger_L\sq_j}(-p',p')
\nonumber\\&&{}
\quad- \GG{\glui_{aR}\epsilonbar\sq_L^\dagger\ysq_j^{\dagger}}(k,-p',p)
\GG{\sq_L\sq^\dagger_j}(p,-p)\Bigr)
\nonumber\\&&{}
- \GG{\sq^\dagger_L\yq\epsilonbar }(-p',p')
\GG{\sq_L\glui_{aR}\qbar}(p,k,-p')
\nonumber\\&&{}
+ \GG{\sq^\dagger_L\glui_{aR} q}(-p',k,p)
\GG{\sq_L\epsilonbar\yqbar}(p,-p)
\nonumber\\&&{}
+ \frac{\delta s\chibar}{\delta\epsilonbar}
\GG{\sq_L\sq_L^\dagger\glui_{aR}\chibar}(p,-p',k,0)
\ .
\label{sqgqSTI}
\end{eqnarray}
In this identity we choose definite interaction eigenstates for the
squarks as external legs and consider only the right-handed part of
the gluino $\glui_R=P_R\glui$. This simplifies the computation, but
due to (only softly broken) $C$- and $P$-invariance it is sufficient
to obtain the supersymmetry-restoring counterterms to all $\sq\glui
q$-interactions.

The identity (\ref{sqgqSTI}) has to hold for arbitrary external momenta. 
Since the counterterms to the interactions are momentum-independent,
it is very convenient to consider the limit $m_i\ll 
|k_\mu| \ll |p_\mu|=|(p'-k)_\mu|$, where $m_i$ denote the masses in
the theory. In this limit all masses can be neglected, and $k$ can be
neglected compared to $p$ except in the terms that would lead to
infrared divergences for $k=0$. The only remaining terms are
\begin{eqnarray}
0 & = &
\GG{\glui_{aR}\epsilonbar \ygluon^{\mu}_c}(k,-k)
\GG{\sq_L\sq_L^\dagger G_{c\mu}}(p,-p',k)
\nonumber\\&&{}
+ \sum_{j=1,2}
\Bigl(\GG{\glui_{aR}\epsilonbar\sq_L\ysq_j}(k,p,-p')
\GG{\sq^\dagger_L\sq_j}(-p',p')
\nonumber\\&&{}
\quad- \GG{\glui_{aR}\epsilonbar\sq_L^\dagger\ysq_j^{\dagger}}(k,-p',p)
\GG{\sq_L\sq^\dagger_j}(p,-p)\Bigr)
\nonumber\\&&{}
- \GG{\sq^\dagger_L\yq\epsilonbar }(-p',p')
\GG{\sq_L\glui_{aR}\qbar}(p,k,-p')
\ .
\end{eqnarray}
The physical meaning of the first and the last term is obvious. They
relate the $\sq\sq G_\mu$ and $\sq q \glui$ interactions, multiplied
with the corresponding supersymmetry transformations of the gluon and
quark, respectively. The other terms are particularly interesting:
They involve supersymmetry transformations of squarks into products of
the  form $\epsilonbar\glui\sq$ --- such transformations are not
present at the tree level but can be generated at one-loop order. The
corresponding Feynman graphs are all finite and thus involve no
regularization ambiguity. In the limit defined above the results are
\begin{eqnarray}
0 & = & -2\psl g P_R T^a\Bigl(1 + \delta_{\ygluon\glui\epsilon}
\nonumber\\&&{}
\qquad\qquad  + \frac{\alpha_s}{4\pi} \CA  
          \big[ B_0(k^2)
                   + B_0 +\frac32 p^2 C_{1}\big]\Bigr)
\nonumber\\&&{}
+\psl g P_R T^a p^2 \frac{\alpha_s}{4\pi}
        \Bigl(\frac{\CA }{2}\big[2C_0+C_{1}\big]
\nonumber\\&&{}
\qquad\qquad + \frac{\CA }{2}\big[2C_0-C_{1}\big] + \CF  \big[2C_{1}\big]\Bigr)
\nonumber\\&&{}
+2\psl g P_R T^a \Bigl(1 + \delta_{\yq\sq\epsilon}+\delta_{\sq\glui q}
\nonumber\\&&{}
\qquad\qquad  +\frac{\alpha_s}{4\pi}
         \big[\CF \big(-B_0 + B_0-p^2 C_{1}\big)
\nonumber\\&&{}
\qquad\qquad   +\frac{\CA }{2}\big(4B_0 + 7p^2 C_{1}
         -2\dreg\big)\big]\Bigr)
\ .
\end{eqnarray}
The arguments of the one-loop functions are as in eq.\ (\ref{BCArgs}),
except where indicated differently. Most of the terms cancel, leaving
only 
\begin{eqnarray}
0 & = & 2\psl g P_R T^a
 \Bigl(-\delta_{\ygluon\glui\epsilon}
       + \frac{\alpha_s}{4\pi}\CA \big[p^2 C_0-B_0(k^2)
\nonumber\\&&{}\qquad
   +2p^2C_{1}+B_0 - 1\dreg\big]
 + \delta_{\yq\sq\epsilon}
 + \delta_{\sq\glui q}\Bigr)
\ .
\end{eqnarray}
For the terms $C_0$ and $B_0(k^2)$, which are infrared divergent for
$|\frac{k}{p}|\to0$, one can easily verify the identity
\begin{eqnarray}
\lim_{k\to0}\big(p^2C_0(p^2,(p+k)^2,k^2,0,0,0)-B_0(k^2,0,0)\big) & = & 
\nonumber\\{}
-2p^2C_{1} - B_0\ .
\end{eqnarray}
Thus, all $B_0,C_i$ functions cancel perfectly, leaving an identity
for the counterterms,
\begin{eqnarray}
0 & = & \delta_{\sq\glui q}
-\delta_{\ygluon\glui\epsilon}-\frac{\alpha_s}{4\pi}\CA \dreg + \delta_{\yq\sq\epsilon}
\ ,
\end{eqnarray}
that has a unique solution for the counterterm to the $\sq\glui q$
interaction
\begin{eqnarray}
 \delta_{\sq\glui q} & = & \frac{\alpha_s}{4\pi}
       \left(\frac23 \CA -\frac12 \CF \right)\dreg
\ ,
\label{sqgluiqRes}
\end{eqnarray}
or written in terms of a counterterm Lagrangian
\begin{eqnarray}
\lefteqn{\L_{{\rm non-sym},\ \sq\glui q}  =  - \delta_{\sq\glui q}
  \sqrt2\,g\,\times}
\nonumber\\&&{}\left(\sq_L^\dagger \gluibar P_L q + \qbar P_R\glui \sq_L
              -\sq_R^\dagger \gluibar P_R q - \qbar P_L\glui \sq_R
         \right)
\ .
\end{eqnarray}
Here the result obtained  for the $\qbar P_R\glui \sq_L$-interaction
has been extended to the other $\sq\glui q$-interactions. The
respective calculations can be done in the same way and yield the same
result due to hermiticity and (softly broken) $C$- and
$P$-invariance. As can be easily checked, $R$-invariance is not
violated by the regularization, and therefore $R$-violating
counterterms like $\qbar P_R \glui \sq_R$ are not necessary.

Hence, in dimensional regularization ($\dreg=1$) an
additional counterterm is necessary to compensate the supersymmetry
breaking of the regularization. In dimensional reduction, however,
this counterterm is not necessary. Both results have already been
anticipated in \cite{MSBar}, so eq.\ (\ref{sqgluiqRes}) provides a
confirmation on the basis of a rigorous definition of the model using
the Slavnov-Taylor identity. 

We want to stress that the result for the counterterm
$\delta_{\sq\glui q}$ is unique. The uniqueness guarantees that
$\delta_{\sq\glui q}$ is not only the solution of (\ref{sqgqSTI}) but
of all symmetry identities. To arrive at this result the unambiguous
calculation of the counterterms to the supersymmetry transformations
in the preceding subsections has been necessary. In particular, in
dimensional reduction only the combination of all these calculations
implies that supersymmetry is preserved in this sector.

\subsection{Gauge interactions}

In the previous subsection we have determined the counterterm to the
$\sq\glui q$ interaction. All other trilinear interactions of
supersymmetric QCD are gauge interactions: $\sq\sq G_\mu$, $qq G_\mu$,
$\glui\glui G_\mu$, and $G_\rho G_\nu G_\mu$. We have explicitely
checked that all symmetry identities that determine these
interactions hold automatically in both regularization schemes,
although $\gamma_5$-interactions are involved. Since these
identities are not due to supersymmetry, and since the gauge
invariance of both schemes is generally known, we are brief in this
subsection and restrict ourselves to a sketch of the calculations.

The identities that determine the counterterms to the gauge
interactions can be obtained from the following derivatives of the
Slavnov-Taylor identity:
\begin{align}
\dfrac{^3 S(\Gamma)}{\sq^\dagger\delta\sq\delta c_a},&&
\dfrac{^4 S(\Gamma)}{\sq \delta c_a\delta c_b \delta\ysq},
\nonumber\\
\dfrac{^3 S(\Gamma)}{q \delta \qbar\delta c_a},&&
\dfrac{^4 S(\Gamma)}{q \delta c_a\delta c_b \delta\yqbar},
\nonumber\\
\dfrac{^3 S(\Gamma)}{\glui_c\delta\gluibar_d\delta c_a},&&
\dfrac{^4 S(\Gamma)}{\glui_c \delta c_a\delta c_b \delta\ygluibar_d},
\nonumber\\
\dfrac{^3 S(\Gamma)}{G_c^\nu\delta G_d^\mu\delta c_a},&&
\dfrac{^4 S(\Gamma)}{c_a\delta c_b \delta\ygluon_d^\mu}.
\label{GaugeIdentities}
\end{align}
The first set of these identities expresses the gauge invariance,
whereas the second set corresponds to the symmetry algebra. Owing to
our parametrization (see sec. \ref{Sec:Parametrization}) there is no
symmetry-violating counterterm to the $\sq\sq G_\mu$
interaction. Rather, this interaction defines the gauge coupling and
determines the other counterterms.\footnote{There is nothing  special
  about this interaction; we could have chosen any other gauge
  interaction instead to define the gauge coupling.} Evaluating these
identities explicitely at 
one-loop order we find that they are satisfied at the regularized
level, which has the following consequence for the
counterterms:\footnote{In particular we find the --- already
  non-trivial --- result that the Lorentz- and SU(3)-structure of the
  counterterms must be identical to the one of the tree-level
  interactions. For instance, in general there could be two linearly
  independent counterterms to the $GGG$ interaction
\begin{eqnarray*}
\L_{\rm non-sym\ }{}_{GGG}& = & \delta_{GGG}\frac12 f_{abc}G^\mu_a G^\nu_b
  \partial_\mu G_c{}_\nu\nonumber\\&&{}
 + \delta_{GGG2}{\rm Tr}(G^\mu G_\mu
  \partial^\nu G_\nu)\ ,
\end{eqnarray*}
but only $\delta_{GGG}$ can contribute.}
\begin{align}
\delta_{c\ygluon}&=\delta_{\sq c\ysq},&
\delta_{\sq c\ysq}&=\delta_{cc\yc},\nonumber\\
\delta_{c\ygluon} + \delta_{qqG}&=\delta_{q c\yq},&
\delta_{q c\yq}&=\delta_{cc\yc},\nonumber\\
\delta_{c\ygluon} + \delta_{\glui\glui G}&=\delta_{\glui c\yglui},&
\delta_{\glui c\yglui}&=\delta_{cc\yc},\nonumber\\
\delta_{c\ygluon} + \delta_{GGG}&=\delta_{c\ygluon G},&
\delta_{c\ygluon G}&=\delta_{cc\yc}.
\end{align}
These identities have the unique solution
\begin{eqnarray}
\delta_{qqG}=\delta_{\glui\glui G}=\delta_{GGG}& = &0\ ,\nonumber\\ 
\delta_{\sq c\ysq}=\delta_{q c\yq}=\delta_{\glui c\yglui}=
\delta_{c\ygluon G}& = & \delta_{c\ygluon}\ ,
\end{eqnarray}
where the only freedom is the value of $\delta_{c\ygluon}$, which is
related to the residue of the ghost propagator and can be fixed by
specifying a renormalization condition for the latter. The important
result, however, is that the non-symmetric counterterms to all gauge
interactions turn out to be zero. To obtain this result, it has
been essential to verify in particular the second set of identities in
eq.\ (\ref{GaugeIdentities}), corresponding to the SU(3)-algebra.

\subsection{Gluon-Gluino-Gluino vertex}

In the previous subsection the result
\begin{eqnarray}
\delta_{\glui\glui G} = \delta_{GGG} & = & 0
\label{ggGandGGGCT}
\end{eqnarray}
was derived using gauge invariance. On the other hand, the $\glui\glui
G_\mu$ and $G_\rho G_\nu G_\mu$ interactions are also related by
supersymmetry. Consistency requires that the relation imposed by
supersymmetry must be automatically satisfied, which gives an
important check. Equivalently, if we use supersymmetry to
rederive the counterterms we must obtain a result compatible with
(\ref{ggGandGGGCT}). 

The following identity, which is due to supersymmetry, connects the
triple-gluon vertex and the gluon-gluino-gluino vertex:
\begin{eqnarray}
0 & = & \frac{\delta^4 S(\Gamma)}{\delta G_a^\mu 
              \delta G_b^\nu \delta\glui_c \delta\epsilonbar}
\nonumber\\
\Rightarrow\quad 0 & = & 
\left[
\GG{G_b^\nu\glui_c\epsilonbar \ygluon_d^\rho}
\GG{G_a^\mu G_d{}_\rho}
+ (_{(\mu,a)\leftrightarrow(\nu,b)})\right]
\nonumber\\&&{}
+
\GG{\glui_c\epsilonbar \ygluon_d^\rho}
\GG{G_a^\mu G_b^\nu G_d{}_\rho}
\nonumber\\&&{}
- \left[
\GG{G_b^\nu\yglui_d\epsilonbar}
\GG{G_a^\mu\glui_c\gluibar_d}
+ (_{(\mu,a)\leftrightarrow(\nu,b)})\right]
\nonumber\\&&{}
-
\GG{G_a^\mu G_b^\nu\yglui_d\epsilonbar}
\GG{\glui_c\gluibar_d}
\nonumber\\&&{}
+\frac{\delta s\chibar}{\delta\epsilonbar}
\GG{G_a^\mu G_b^\nu \glui_c \chibar}
\ .
\label{GGGGggSTI}
\end{eqnarray}
The vertex functions involving the $\epsilon$ ghost correspond to
supersymmetry transformations. 

The identity (\ref{GGGGggSTI}) has to hold for arbitrary external
momenta, but for the purpose of deriving counterterms it is sufficient
to look at the limit $m_i\ll |p_b| \ll |p_a|$ and to consider only the
leading terms in $p_a$. In this limit the soft-breaking term and all
the masses do not contribute, and all contributions get a simple
analytical form, and all non-local terms cancel:
\begin{eqnarray}
0 & = &{}i g f_{abc}\bigg(\Big(\frac{2\CA}{3}
\frac{\alpha_s}{4\pi}\dreg
\nonumber\\&&{}
 + \delta_{GGG} +
          \delta_{\ygluon\glui\epsilon} - \delta_{\yglui\epsilon GG}\Big)
            \psl_a g_{\mu\nu}
\nonumber\\&&{}
  + \Big(
          - \delta_{G\glui\glui} -
          \delta_{G\epsilon\yglui} + \delta_{\yglui\epsilon GG}\Big)
            \gamma_\mu\gamma_\nu \psl_a
\nonumber\\&&{}
  + \Big(\frac{2\CA}{3} \frac{\alpha_s}{4\pi}\dreg
         - \delta_{G\glui\glui} + \delta_{GGG}
\nonumber\\&&{}
         - \delta_{G\epsilon\yglui} + \delta_{\ygluon\glui\epsilon}\Big)
    (\gamma_\nu p_a{}_\mu - 2 \gamma_\mu p_a{}_\nu)\bigg)
\label{ResultGggSTI}
\ .
\end{eqnarray}
From the longitudinal part of this equation, which is obtained by the
contraction with $p_a^\mu$, we get
\begin{eqnarray}
\delta_{\yglui\epsilon GG} & = & \frac{2\CA}{3} \frac{\alpha_s}{4\pi}\dreg
                            + \delta_{GGG} +
                              \delta_{\ygluon\glui\epsilon}
\ .
\end{eqnarray}
This result fixes the counterterm for the vertex function $\GG{G_a^\mu
  G_b^\nu\yglui_d\epsilonbar}$ in terms of the other counterterms.
Inserting it in (\ref{ResultGggSTI}) yields
\begin{eqnarray}
\delta_{G\glui\glui} & = & \frac{2\CA}{3} \frac{\alpha_s}{4\pi}\dreg + \delta_{GGG} -
\delta_{G\epsilon\yglui} + 
\delta_{\ygluon\glui\epsilon}
\ .
\label{Result2GggSTI}
\end{eqnarray}
This is an expression for the counterterm of the gluon-gluino-gluino
interaction in terms of the counterterm of the triple-gluon vertex and
the counterterms $\delta_{G\epsilon\yglui}$ and
$\delta_{\ygluon\glui\epsilon}$, calculated in subsec.\
\ref{Sec:GgSelfEnergies}. The results for $\delta_{G\epsilon\yglui}$
and $\delta_{\ygluon\glui\epsilon}$ show that these counterterms
cancel the first term on the r.h.s.\ of eq.\ (\ref{Result2GggSTI}),
thus yielding
\begin{eqnarray}
\delta_{G\glui\glui} & = & \delta_{GGG}
\ ,
\end{eqnarray}
in agreement with eq.\ (\ref{ggGandGGGCT}).

\section{Conclusions}
\label{Sec:Conclusions}

We have calculated symmetry-restoring counterterms in supersymmetric
QCD. We have found that in DRed no non-symmetric counterterms are
necessary at the one-loop level, neither for the trilinear
interactions nor for the symmetry transformation operators. While
identities like eq.\ (\ref{GgSTI}) have been checked in the literature
\cite{CJN80}, the result that the loop-corrected symmetry
transformations automatically satisfy the right algebra is new. In
DReg counterterms for the $\sq\glui q$-interaction and for most
symmetry transformations are required.

In order to obtain unique results we have had to take into
account loop corrections to the symmetry transformations and to
renormalize them in such a way that the SU(3)- and the supersymmetry
algebra are satisfied. The fact that the counterterms to these
symmetry transformations can be unambiguously calculated is the main
advantage of the Slavnov-Taylor identity compared to simple
supersymmetric Ward identities like eq.\ (\ref{SusyWI}). 

The uniqueness of our results guarantees that the counterterms derived
from only few symmetry identities remain correct even if we take into
account all symmetry identities simultaneously. Only to the vertex
functions we have not considered additional counterterms may be
required. Therefore our results for DRed, where no counterterm is
needed, constitute a definite test of the supersymmetry of the
scheme for the considered cases. And when the use of dimensional
regularization is desirable, for instance because the standard
$\overline{MS}$ mass-factorization scheme should be used
\cite{BHSZ96}, our results show how to establish all symmetries.

Let us now give some final remarks on the necessity to calculate loop
corrections and counterterms to the symmetry transformations. This
necessity might seem disturbing. But using the Feynman
rules and the explicit form of the Slavnov-Taylor identity we have
provided, the calculations turn out to be straightforward. In fact,
the one-loop corrections to the symmetry transformations are much
simpler than the one-loop corrections to the interaction vertices we
have considered. 

\begin{figure}[htb]
\begin{center}
\begin{picture}(415,69)
\allpsfrag
\epsfxsize=8cm
\put(0,-80){\epsfbox{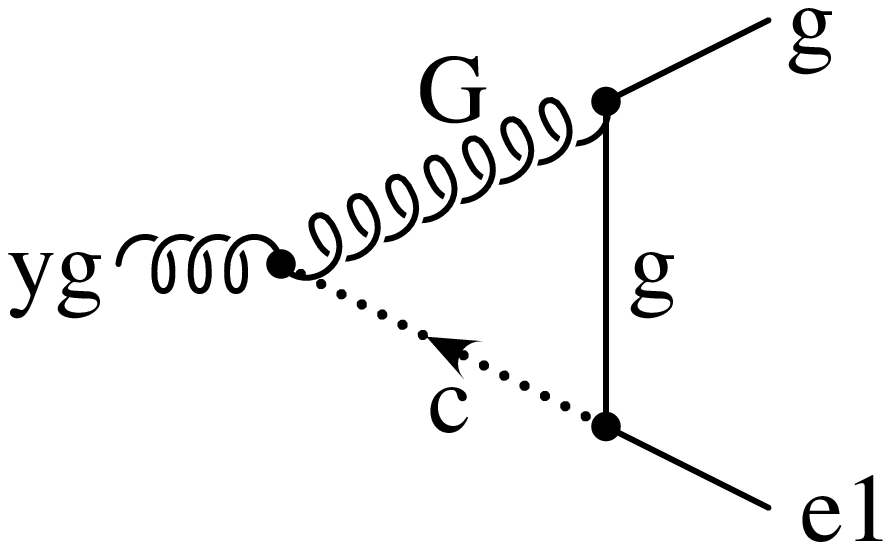}}
\end{picture}
\end{center}
\caption{The one-loop diagram contributing to the vertex function
  $\GG{\ygluon^\mu \glui\epsilonbar}$.}
\label{FigEpsYg}
\end{figure}

As mentioned in \cite{HKS99} the appearance of loop corrections to the
symmetry transformations can be traced back to two reasons. First, the
non-linearity of the BRS transformations can cause a difference
between expectation values $\langle s\varphi\rangle$ and the respective
products of the classical fields. In the loop diagrams the
non-linearity is the reason for the triple- or quartic couplings to
the external $Y$ sources. A second reason is the
supersymmetry breaking of the gauge fixing, which necessitates
compensating terms involving the supersymmetry ghosts. The
corresponding Feynman rules appear in most of the loop diagrams to the
supersymmetry transformations. Owing to these terms, loop
corrections are even possible to supersymmetry transformations that
are linear at the tree level, e.g.\ the one of the gluon (see fig.\
\ref{FigEpsYg}). 

This sheds a light on the deep connection of gauge
invariance and supersymmetry, which was a major complication in
the renormalization for a long time and has enforced the introduction
of the Slavnov-Taylor identity. Our results help to get a quantitative
understanding of such general properties of the theory.

\acknowledgement{We thank E.\ Kraus for helpful comments and T.\
  Fritzsche, T.\ Hahn, and C.\ Schappacher for useful discussions and
  valuable advice in the use of the {\em Mathematica} packages {\em
  FeynArts} and {\em FormCalc}.}

\begin{appendix}

\section{Lagrangian and BRS transformations}
\label{Sec:Construction}

In this section we give the explicit form of the BRS transformations
and the Lagrangian of supersymmetric QCD as a specialized version of
the general Yang-Mills theories discussed in \cite{SYM}. In contrast
to there, we use 4-spinors throughout in order to obtain Feynman rules
that can be used in a straightforward way. 

\subsection{BRS Transformations}
\label{Sec:BRS}

We combine gauge and supersymmetry
transformations and translations in a single anticommuting BRS
operator $s$. On the ``physical'' fields (i.e.\ the ones carrying no
ghost number) $s$ acts as the sum of gauge and
supersymmetry transformations and translations, where the
transformation parameters have been promoted to ghost fields $c_a(x),
\epsilon, \omega^\mu$:
\begin{eqnarray}
sG^\mu & = & \partial^\mu c - ig[c,G^\mu] 
           + \epsilonbar\gamma^\mu\glui
- i\omega^\nu\partial_\nu G^\mu
\ ,\nonumber\\
s\glui & = & -ig\{c,\glui\}
           - \frac12\sigma^{\rho\sigma}\epsilon F_{\rho\sigma}
           + D (P_L-P_R) \epsilon
- i\omega^\nu\partial_\nu\glui
\ ,\nonumber\\
s\gluibar & = & -ig\{c,\gluibar\}
           + \frac12\epsilonbar\sigma^{\rho\sigma} F_{\rho\sigma}
           + \epsilonbar (P_L-P_R)D
- i\omega^\nu\partial_\nu\gluibar
\ ,\nonumber\\
s\sq_L & = & -igc\sq_L + \sqrt2\epsilonbar P_L q -
i\omega^\nu\partial_\nu\sq_L
\ ,\nonumber\\
s\sq_L^\dagger & = & +ig\sq_L^\dagger c + \sqrt2\qbar P_R \epsilon -
i\omega^\nu\partial_\nu\sq_L^\dagger
\ ,\nonumber\\
s\sq_R & = & -igc\sq_R - \sqrt2\epsilonbar P_R q -
i\omega^\nu\partial_\nu\sq_R
\ ,\nonumber\\
s\sq_R^\dagger & = & +ig\sq_R^\dagger c - \sqrt2\qbar P_L \epsilon -
i\omega^\nu\partial_\nu\sq_R^\dagger
\ ,\nonumber\\
sq & = & -igcq + \sqrt2 m(\sq_L P_R - \sq_R P_L)\epsilon
\nonumber\nonumber\\&&{}
         + \sqrt2 i D_\mu(\sq_L P_L - \sq_R P_R)\gamma^\mu\epsilon -
i\omega^\nu\partial_\nu q
\ ,\nonumber\\
s\qbar & = & -ig\qbar c
         + \sqrt2 m\epsilonbar(-\sq_L^\dagger P_L + \sq_R^\dagger P_R)
\nonumber\\&&{}
         + \sqrt2 i \epsilonbar \gamma^\mu
                 ((D_\mu\sq_L)^\dagger P_R - (D_\mu\sq_R)^\dagger P_L) -
i\omega^\nu\partial_\nu \qbar\ ,\nonumber\\
sa & = & \sqrt2\epsilonbar P_L \chi - i\omega^\nu\partial_\nu a\ ,\nonumber\\
sa^\dagger & = & \sqrt2\chibar P_R \epsilon 
               - i\omega^\nu\partial_\nu a^\dagger\ ,\nonumber\\
s\chi & = & \sqrt2 (P_L \hat{f}- P_R\hat{f}^\dagger)\epsilon
\nonumber\\&&{}
          + \sqrt2 i (P_L\partial_\mu a - P_R\partial_\mu
          a^\dagger)\gamma^\mu\epsilon -i\omega^\nu\partial_\nu\chi \ ,\nonumber\\
sf & = & \sqrt2 i \epsilonbar\gamma^\mu\partial_\mu P_L\chi
         - i\omega^\nu\partial_\nu f\ ,\nonumber\\
sf^\dagger & = & -\sqrt2 i \partial_\mu\chibar\gamma^\mu P_L\epsilon 
         - i\omega^\nu\partial_\nu f^\dagger\ .
\end{eqnarray}
Here we have used the notation $G^\mu=G_a^\mu T^a$ etc.\ for all
fields in the adjoint representation of the gauge group. Furthermore,
we have used the gauge covariant derivative 
\begin{eqnarray}
D^\mu & = & \partial^\mu +igT^aG_a^\mu\ ,
\end{eqnarray}
where $T^a$ has to be replaced by $-if_{abc}$ in the adjoint
representation, the field strength tensor
\begin{eqnarray}
F_a^{\mu\nu} & = & \partial^\mu G_a^\nu-\partial^\nu G_a^\mu
                   - g f_{abc} G^\mu_b G^\nu_c\ ,
\end{eqnarray}
and the abbreviation
\begin{eqnarray}
D_a & = & -g(\sq_L^\dagger T^a \sq_L - \sq_R^\dagger T^a \sq_R)\ .
\label{DDefinition}
\end{eqnarray}

Generally, the BRS operator has the important nilpotency property
\begin{eqnarray}
s^2 & = & 0 + \mbox{field equations}
\label{Nilpotency}
\end{eqnarray}
provided the statistics of the ghost fields is ``wrong'', i.e.\
opposite to the one required by the spin-statistics theorem, and the
BRS transformations of the ghosts themselves are given by the
structure constants of the symmetry algebra, as follows:
\begin{eqnarray}
sc & = & -igc^2 + i\epsilonbar\gamma^\mu\epsilon G_\mu
- i\omega^\nu\partial_\nu c
\ ,\\
s\epsilon & = & 0
\ ,\\
s\omega^\nu & = & \epsilonbar\gamma^\nu\epsilon
\ .
\end{eqnarray}
Finally, for gauge fixing we introduce Faddeev-Popov antighosts
$\cbar_a$ and auxiliary fields $B_a$ with BRS transformations
\begin{eqnarray}
s\cbar & = & B - i\omega^\nu\partial_\nu \cbar
\ ,\\
sB & = & i\epsilonbar\gamma^\nu\epsilon \partial_\nu \cbar
- i\omega^\nu\partial_\nu B\ .
\end{eqnarray}

\subsection{Lagrangian}
\label{Sec:L}
The Lagrangian of supersymmetric QCD consists of one part containing
the physical fields only, and one part containing the ghosts and the
external fields. The first part is given by
\begin{eqnarray}
\lefteqn{\L_{\rm phys}  =  \L_{\rm kin} + \L_m + \L_{\rm soft}\ ,}
\nonumber\\
\lefteqn{\L_{\rm kin}  =  
-\frac{1}{4}(F_{\mu\nu}^a)^2
+ \frac12 \gluibar_a i\gamma^\mu D_\mu \glui_a}
\nonumber\\&&{}
+\qbar i\gamma^\mu D_\mu q
 + |D^\mu\sq_L|^2 + |D^\mu\sq_R|^2
 - \frac12 D_a D_a
\nonumber\\&&{}
 -\sqrt2g\left(\sq_L^\dagger \gluibar P_L q + \qbar P_R\glui \sq_L
              -\sq_R^\dagger \gluibar P_R q - \qbar P_L\glui \sq_R
         \right)
\ ,\nonumber\\
\lefteqn{\L_{m}  =  -m\qbar q -m^2\left(|\sq_L|^2+|\sq_R|^2\right)
\ ,}\nonumber\\
\lefteqn{\L_{\rm soft}  =  -\frac12\tilde{m}{}_{\glui} \,
                     \gluibar\, (P_L \hat{f}+P_R
                     \hat{f}^\dagger)\,\glui
+ {\cal O}(a,a^\dagger,\chi)
}
\nonumber\\&&{}
-\left(\tilde{q}_L^\dagger\ \tilde{q}_R^\dagger\right)
 \left(\begin{array}{rr}
       |\hat{f}|^2\tilde{M}_L^2  &  m\hat{f}^\dagger\tilde{M}_{LR} \\ 
       m\hat{f}\tilde{M}_{LR}    &  |\hat{f}|^2\tilde{M}_R^2 
       \end{array}\right)
\left({\begin{array}{c}\tilde{q}_L \\
    \tilde{q}_R\end{array}}\right) 
\ .
\label{LSoft}
\end{eqnarray}
In the soft-breaking terms we have not written out the explicit form
of the terms containing the $a$, $\chi$ components of the chiral
supermultiplet we use to generate the soft breaking. The usual
breaking terms are obtained by setting $\hat{f}$ to the constant
$f_0$.

The parts of the Lagrangian containing the ghosts and the external
fields are $\L_{\rm fix,\ gh}$ as given in eq.\ (\ref{GaugeFixing}) and
\begin{eqnarray}
\L_{\rm ext} & = & \ygluon^\mu_a sG_{a\mu} + \ygluibar s\glui
+ \yc sc
\nonumber\\&&{}
 + \ysq_L s\sq_L
 + (s\sq_L^\dagger)\ysq_L^\dagger
 + \ysq_R s\sq_R
\nonumber\\&&{}
 + (s\sq_R^\dagger)\ysq_R^\dagger
 + \yqbar s q - (s\qbar)\yq\ ,
\\
\L_{\rm bil} & = & \frac12(\ygluibar_a(P_R-P_L)\epsilon)
                          (\epsilonbar(P_R-P_L)\yglui_a)
\nonumber\\&&{}   -2(\yqbar P_R \epsilon)(\epsilonbar P_L \yq)
                  -2(\yqbar P_L \epsilon)(\epsilonbar P_R \yq)
\ .
\end{eqnarray}
Then the classical action is the sum of these parts:
\begin{eqnarray}
\Gamma_{\rm cl} & = & \intx \left(\L_{\rm kin}+\L_{m}+\L_{\rm
  soft}+\L_{\rm fix,\ gh}+\L_{\rm ext}+\L_{\rm bil}\right)
\nonumber\\&&{}
+{\cal O}(a,a^\dagger,\chi)\ .
\end{eqnarray}
It satisfies all defining symmetry requirements of sec.\
\ref{Sec:Definition}.

\section{Feynman rules}
\label{Sec:FR}

In this section we give a list of the Feynman rules we need in our
calculations, in particular of the ones involving the external $Y$
fields and the $\epsilon$ ghosts. 
\begin{itemize}
\item We take all momenta as
incoming. 
\item The $\epsilon$ ghosts are space-time independent
constants and do not carry a momentum. 
\item Many of the following Feynman
rules have to be used with different orderings of the fermions. In the
case of fermionic spinors the flipping rules of \cite{DEHK92} have to
be applied, and in the case of fermionic scalars and vectors the
orderings correspond to different global signs of the vertices. For
ease of reference we give the alternative rules explicitely for the
cases we will need later.
\end{itemize}
For brevity, we write $\Gamma$ instead of $\Gamma_{\rm cl}$ for the
classical action in this section.

\newcommand{\frwidth}{3.5cm}
\newcommand{\frpicheight}{45}
\newcommand{\frpicwidth}{100}
\newcommand{\frputleft}{-30}
\begin{align*}
\parbox[b]{\frwidth}
{
\begin{picture}(\frpicwidth,\frpicheight)
\psfrag{e1}{$G$}
\psfrag{e2}{$G$}
\psfrag{e3}{$G$}
\epsfxsize=6cm
\put(\frputleft,-104){\epsfbox{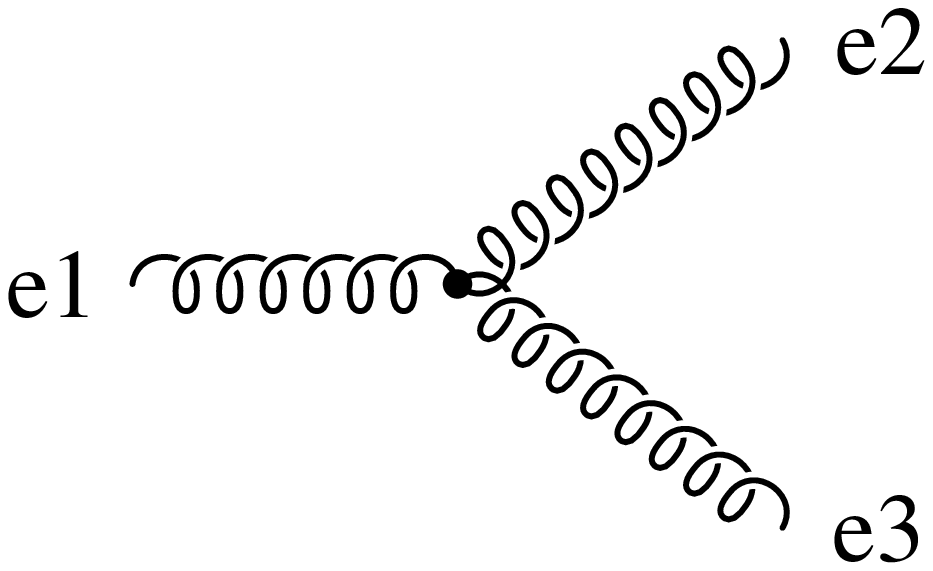}}
\end{picture}
}
&&&
i\GG{G^\rho_a G^\mu_b G^\nu_c}(p_a,p_b,p_c) \nonumber\\*[1ex]&&{}={}&  -gf_{abc}
        [g_{\rho\mu}(p_a-p_b)_\nu 
\nonumber\\* &&&{}
        \quad+g_{\mu\nu}(p_b-p_c)_\rho
\nonumber\\* &&&{}
        \quad+g_{\nu\rho}(p_c-p_a)_\mu]\\[2ex]
\parbox[b]{\frwidth}
{
\begin{picture}(\frpicwidth,\frpicheight)
\psfrag{e1}{$G$}
\psfrag{e2}{$G$}
\psfrag{e3}{$G$}
\psfrag{e4}{$G$}
\epsfxsize=6cm
\put(\frputleft,-104){\epsfbox{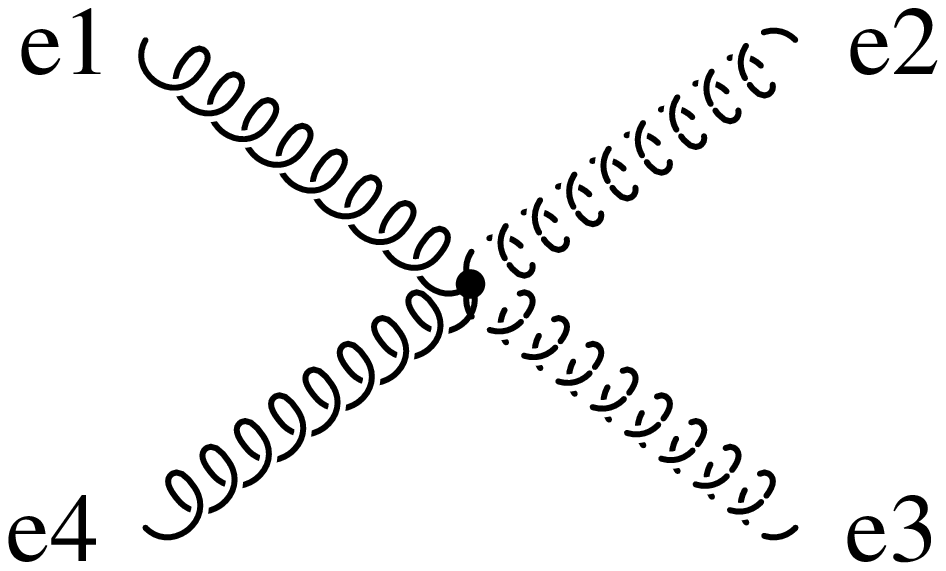}}
\end{picture}
}
&&&
i\GG{G^\mu_a G^\nu_b G^\rho_e G^\sigma_f} \nonumber\\*[1ex]&&{}={}&
-ig^2
\big[
f_{abc}f_{efc}(g_{\mu\rho}g_{\sigma\nu}-g_{\mu\sigma}g_{\nu\rho})
\nonumber\\* &&&
+ f_{aec}f_{fbc}(g_{\mu\sigma}g_{\nu\rho}-g_{\mu\nu}g_{\rho\sigma})
\nonumber\\* &&&{}
+ f_{afc}f_{bec}(g_{\mu\nu}g_{\rho\sigma}-g_{\mu\rho}g_{\sigma\nu})
\big]
\\[2ex]
\parbox[b]{\frwidth}
{
\begin{picture}(\frpicwidth,\frpicheight)
\psfrag{e1}{$G$}
\psfrag{e2}{$\gluibar$}
\psfrag{e3}{$\glui$}
\epsfxsize=6cm
\put(\frputleft,-104){\epsfbox{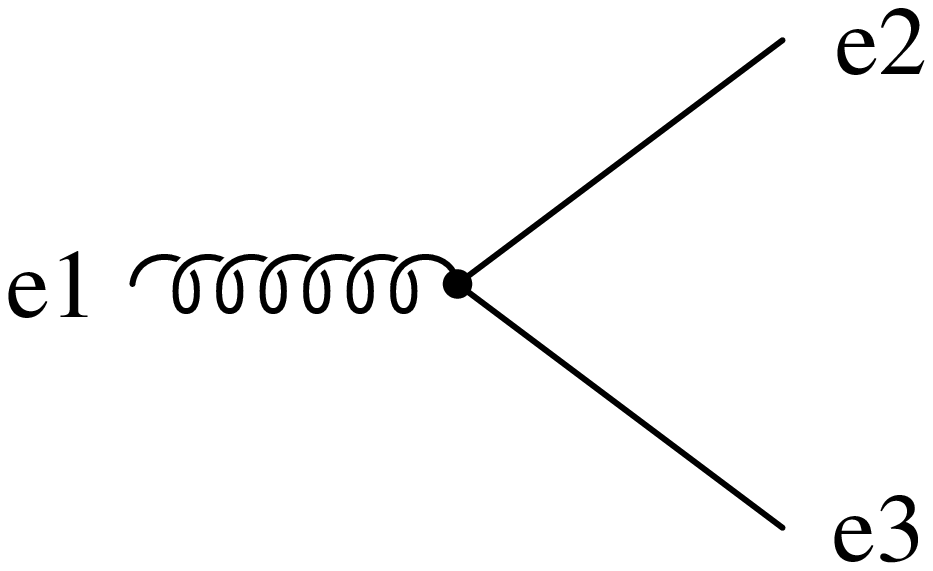}}
\end{picture}
}
&&&
i\GG{\glui_c G^\nu_e \gluibar_d} \nonumber\\*[1ex]&&{}={}&  -gf_{edc}\gamma_\nu\\[2ex]
\parbox[b]{\frwidth}
{
\begin{picture}(\frpicwidth,\frpicheight)
\psfrag{e1}{$G$}
\psfrag{e2}{$\qbar$}
\psfrag{e3}{$q$}
\epsfxsize=6cm
\put(\frputleft,-104){\epsfbox{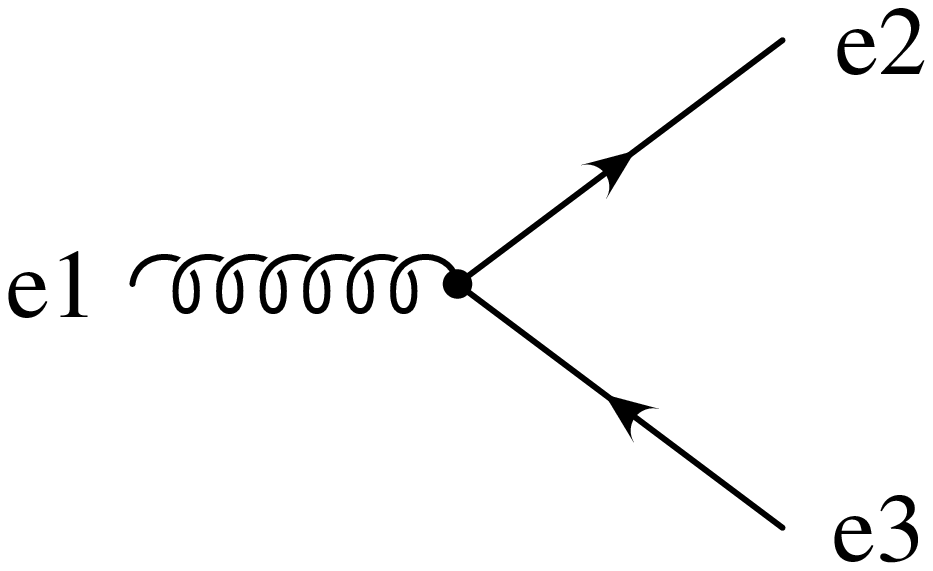}}
\end{picture}
}
&&&
i\GG{G_a^\mu q_j \qbar_i} \nonumber\\*[1ex]&&{}={}& -ig\gamma_\mu T^a_{ij}\\[2ex]
\parbox[b]{\frwidth}
{
\begin{picture}(\frpicwidth,\frpicheight)
\psfrag{e1}{$\hspace{-3ex}G(k)$}
\psfrag{e2}{\hspace{-4ex}\parbox{7ex}{$\sq^\dagger(-p')\\[2.5ex]\mbox{\ }$}}
\psfrag{e3}{$\sq(p)$}
\epsfxsize=6cm
\put(\frputleft,-104){\epsfbox{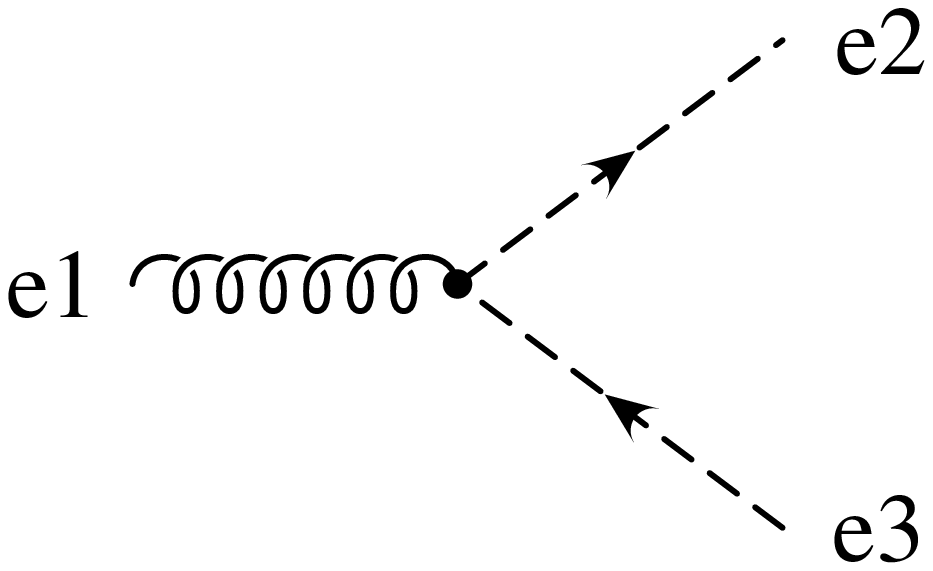}}
\end{picture}
}
&&&
i\GG{G_a^\mu \sq_{k,j} \sq^\dagger_{l,i}}(k, p, -p')
 \nonumber\\*[1ex]&&{}={}& -ig(p+p')_\mu \delta_{kl}T^a_{ij}\\[2ex]
\parbox[b]{\frwidth}
{
\begin{picture}(\frpicwidth,\frpicheight)
\psfrag{e1}{$\glui$}
\psfrag{e2}{$\qbar$}
\psfrag{e3}{$\sq$}
\epsfxsize=6cm
\put(\frputleft,-104){\epsfbox{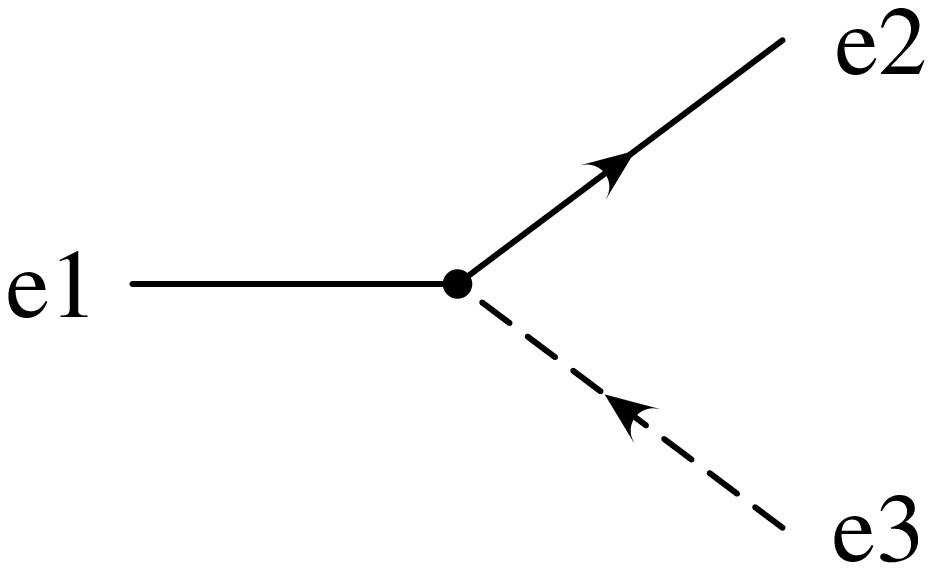}}
\end{picture}
}
&&&
i\GG{\sq_{k,j}\glui_a\qbar_i} \nonumber\\*[1ex]&&{}={}& -i\sqrt2 g T^a_{ij}
\times\nonumber\\* &&&{} (P_RS^*_{kL} - P_LS^*_{kR}) \\[2ex]
\parbox[b]{\frwidth}
{
\begin{picture}(\frpicwidth,\frpicheight)
\psfrag{e1}{$\gluibar$}
\psfrag{e2}{$\sq^\dagger$}
\psfrag{e3}{$q$}
\epsfxsize=6cm
\put(\frputleft,-104){\epsfbox{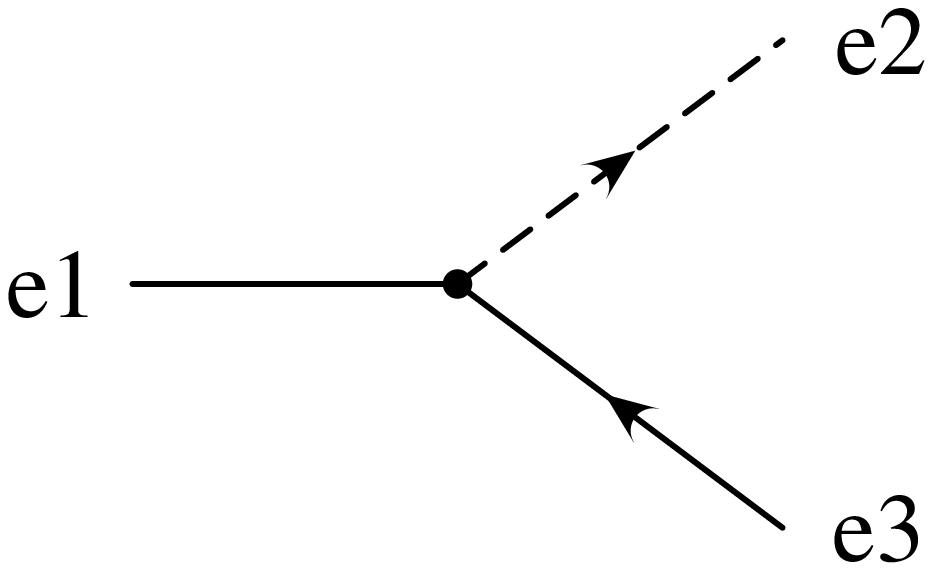}}
\end{picture}
}
&&&
i\GG{q_j \gluibar_a \sq^\dagger_{k,i}} \nonumber\\*[1ex]&&{}={}& -i\sqrt2 g T^a_{ij}
\times\nonumber\\* &&&{} (P_LS_{kL} - P_RS_{kR})\\[2ex]
\parbox[b]{\frwidth}
{
\begin{picture}(\frpicwidth,\frpicheight)
\psfrag{e1}{$G$}
\psfrag{e2}{$\sq^\dagger$}
\psfrag{e3}{$\sq$}
\psfrag{e4}{$G$}
\epsfxsize=6cm
\put(\frputleft,-104){\epsfbox{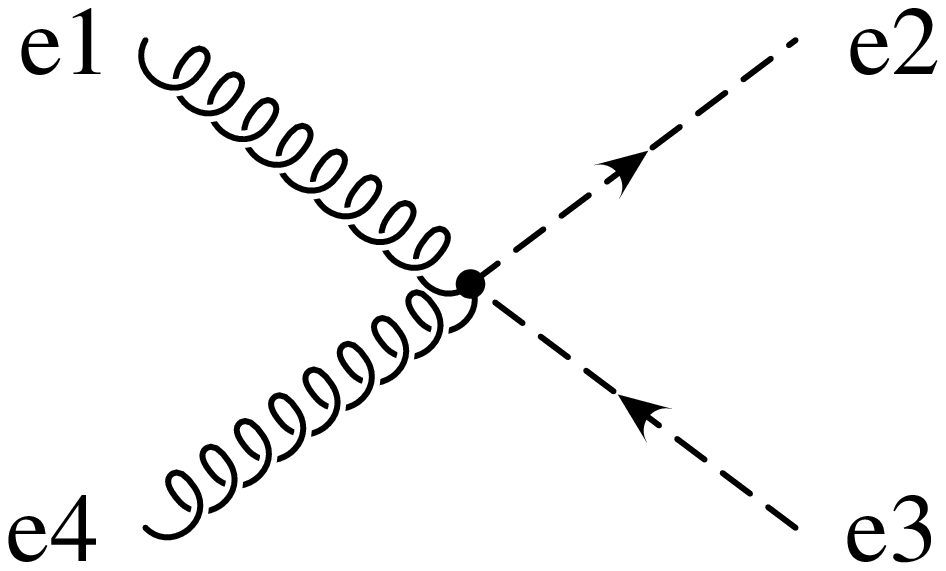}}
\end{picture}
}
&&&
i\GG{G^\mu_a G^\nu_b \sq_{k,j} \sq^\dagger_{l,i}} \nonumber\\*[1ex]&&{}={}& ig^2 g_{\mu\nu}
 \delta_{kl}\{T^a, T^b\}_{ij}\\[2ex]
\parbox[b]{\frwidth}
{
\begin{picture}(\frpicwidth,\frpicheight)
\psfrag{e1}{$G$}
\psfrag{e2}{\parbox{7ex}{$\cbar(-k)$}\\[2.5ex]\mbox{\ }}
\psfrag{e3}{$c$}
\epsfxsize=6cm
\put(\frputleft,-104){\epsfbox{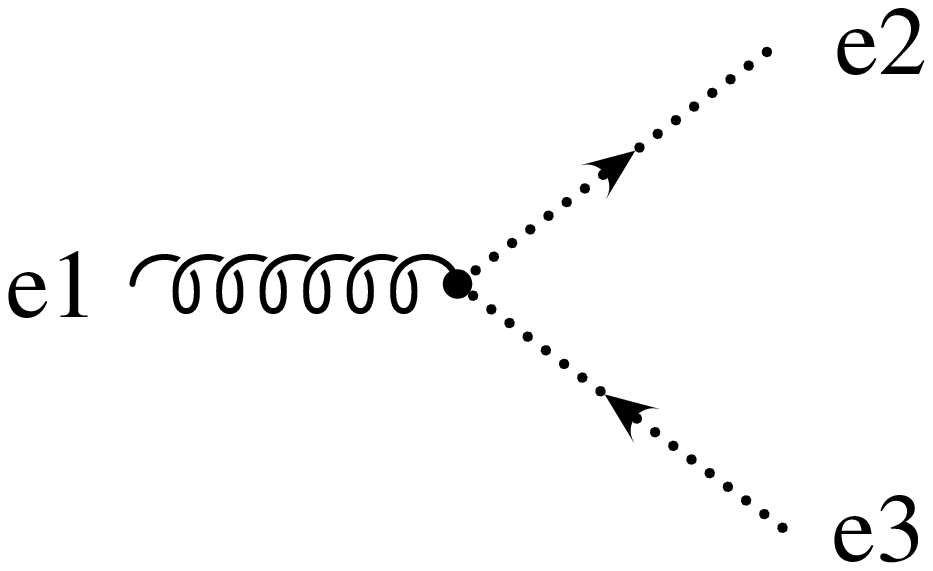}}
\end{picture}
}
&&&
i\GG{c_c G^\mu_b \cbar_a}(k_1,k_2,-k) \nonumber\\*[1ex]&&{}={}&  igf_{bac}ik_\mu
\\[2ex]
\parbox[b]{\frwidth}
{
\begin{picture}(\frpicwidth,\frpicheight)
\psfrag{e1}{\parbox{7ex}{$\glui(k)$\\[2.5ex]\mbox{\ }}}
\psfrag{e2}{$\epsilonbar$}
\psfrag{e3}{$\cbar(-k)$}
\epsfxsize=6cm
\put(\frputleft,-104){\epsfbox{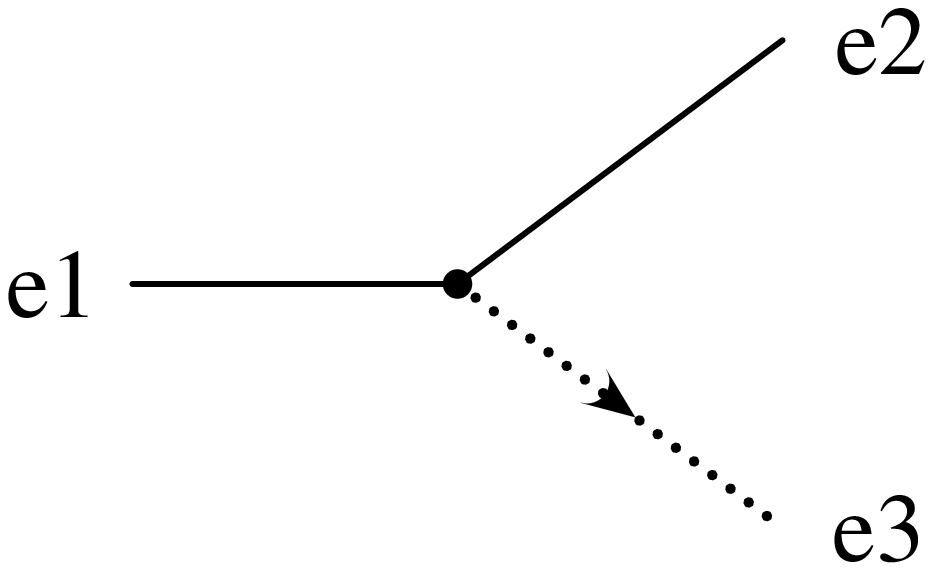}}
\end{picture}
}
&&&
i\GG{\glui_c\epsilonbar\cbar_b}(k,-k) \nonumber\\*[1ex]&&{}={}&  -\ksl\delta_{bc}\\[2ex]
\parbox[b]{\frwidth}
{
\begin{picture}(\frpicwidth,\frpicheight)
\psfrag{e1}{\parbox{7ex}{$\gluibar(-k)$\\[2.5ex]\mbox{\ }}}
\psfrag{e2}{$\epsilon$}
\psfrag{e3}{$\cbar(k)$}
\epsfxsize=6cm
\put(\frputleft,-104){\epsfbox{diagrams/ggcb.ps}}
\end{picture}
}
&&&
i\GG{\cbar_b\epsilon\gluibar_c}(k,-k) \nonumber\\*[1ex]&&{}={}&  -\ksl\delta_{bc}\\[2ex]
\parbox[b]{\frwidth}
{
\begin{picture}(\frpicwidth,\frpicheight)
\psfrag{e1}{\ $\cbar$}
\psfrag{e4}{\ $\cbar$}
\psfrag{e3}{$\epsilonbar$}
\psfrag{e2}{$\epsilon$}
\epsfxsize=6cm
\put(\frputleft,-104){\epsfbox{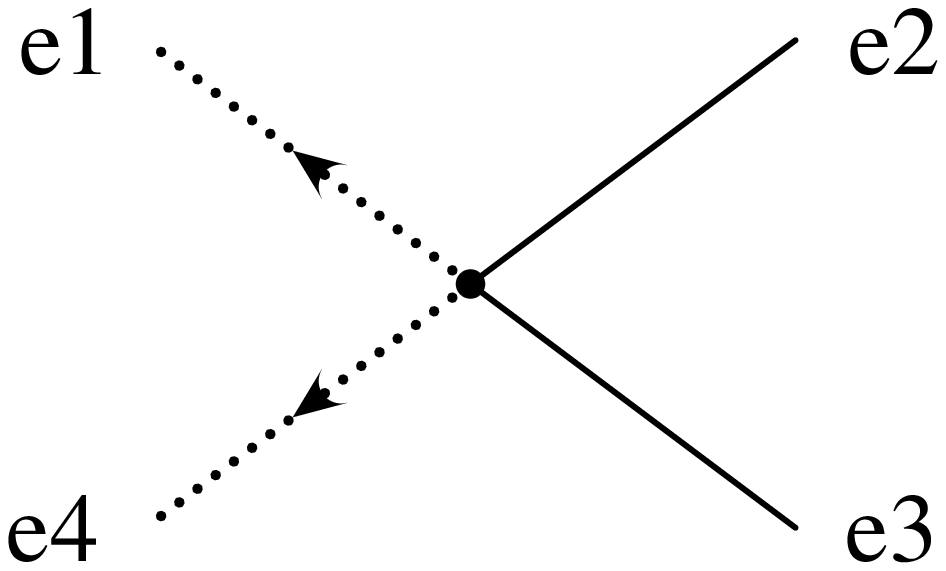}}
\end{picture}
}
&&&
i\GG{\epsilon\epsilonbar\cbar_b\cbar_a}(-k,k) \nonumber\\*[1ex]&&{}={}&
 2i\xi\ksl\delta_{ab}
\\[2ex]
\parbox[b]{\frwidth}
{
\begin{picture}(\frpicwidth,\frpicheight)
\psfrag{e1}{$\hspace{-1ex}\ygluon$}
\psfrag{e2}{$c$}
\psfrag{e3}{$G$}
\epsfxsize=6cm
\put(\frputleft,-120){\epsfbox{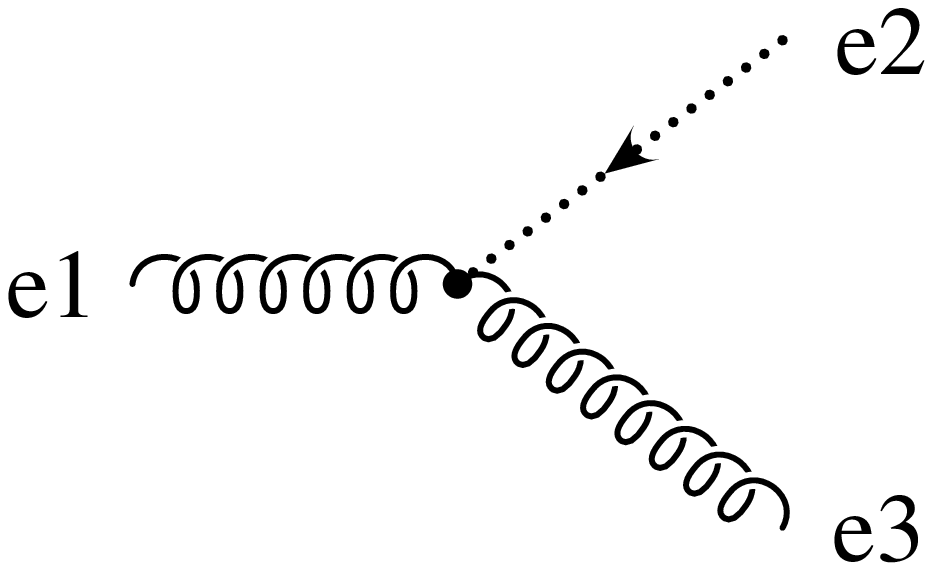}}
\end{picture}
}
&&&i\GG{\ygluon^{\mu}_a c_b G^\nu_c} \nonumber\\*[1ex]&&{}={}&  -igf_{abc}
\\*[2ex]
&&&
i\GG{c_b G^\nu_c \ygluon^{\mu}_a} \nonumber\\*[1ex]&&{}={}&  igf_{abc}g_{\mu\nu}\\[2ex]
\parbox[b]{\frwidth}
{
\begin{picture}(\frpicwidth,\frpicheight)
\psfrag{e1}{$\hspace{-1ex}\ygluon$}
\psfrag{e2}{\hspace{-3ex}\parbox{7ex}{$\{\gluibar,\glui\}$\\[2.5ex]\mbox{\ }}}
\psfrag{e3}{\hspace{-3ex}\parbox{7ex}{\mbox{\ }\\[2.5ex]$\{\epsilon,\epsilonbar\}$}}
\epsfxsize=6cm
\put(\frputleft,-120){\epsfbox{diagrams/Ggg.ps}}
\end{picture}
}
&&&
i\GG{\ygluon{}^\mu_a\epsilon\gluibar_c} \nonumber\\*[1ex]&&{}={}& -i\gamma_\mu\delta_{ac}\\*[2ex]
&&&
i\GG{\ygluon{}^\mu_a\glui_c\epsilonbar} \nonumber\\*[1ex]&&{}={}& -i\gamma_\mu\delta_{ac}\\[2ex]
\parbox[b]{\frwidth}
{
\begin{picture}(\frpicwidth,\frpicheight)
\psfrag{e1}{\parbox{7ex}{$\{\yglui,\ygluibar\}$\\[2.5ex]\mbox{\ }}}
\psfrag{e2}{\hspace{-3ex}\parbox{7ex}{$\{\gluibar,\glui\}$\\[2.5ex]\mbox{\ }}}
\psfrag{e3}{$c$}
\epsfxsize=6cm
\put(\frputleft,-120){\epsfbox{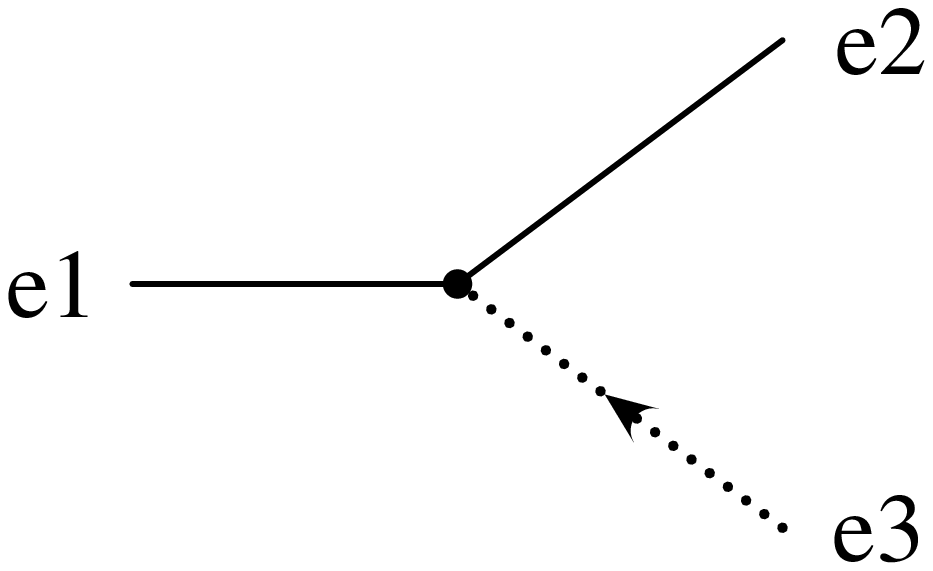}}
\end{picture}
}
&&&
i\GG{\yglui_c c_e \gluibar_d} \nonumber\\*[1ex]&&{}={}&  igf_{ced}\\*[2ex]
&&&
i\GG{\glui_d c_e \ygluibar_c} \nonumber\\*[1ex]&&{}={}&  igf_{ced}\\[2ex]
\parbox[b]{\frwidth}
{
\begin{picture}(\frpicwidth,\frpicheight)
\psfrag{e1}{$\hspace{-5ex}G(-q)$}
\psfrag{e2}{\hspace{-3ex}\parbox{7ex}{$\{\epsilonbar,\epsilon\}\\[2.5ex]\mbox{\ }$}}
\psfrag{e3}{\hspace{-14ex}\parbox{7ex}{$\mbox{\ }\\[2.5ex]\{\yglui(q),\ygluibar(q)\}$}}
\epsfxsize=6cm
\put(\frputleft,-120){\epsfbox{diagrams/Ggg.ps}}
\end{picture}
}
&&&
i\GG{G^\nu_b\yglui_c\epsilonbar}(-q,q) \nonumber\\*[1ex]&&{}={}&
-\sigma_{\nu\mu}q^\mu\delta_{bc} \\*[2ex]
&&&
i\GG{G^\nu_b\epsilon\ygluibar_c}(-q,q) \nonumber\\*[1ex]&&{}={}&
-\sigma_{\nu\mu}q^\mu\delta_{bc}\\[2ex]
\parbox[b]{\frwidth}
{
\begin{picture}(\frpicwidth,\frpicheight)
\psfrag{e1}{$G$}
\psfrag{e2}{\parbox{7ex}{\hspace{-4ex}$\{\ygluibar,\yglui\}\\[2ex]\mbox{\ }$}}
\psfrag{e3}{\hspace{-3ex}\parbox{7ex}{\mbox{\ }\\[2.5ex]$\{\epsilon,\epsilonbar\}$}}
\psfrag{e4}{$G$}
\epsfxsize=6cm
\put(\frputleft,-120){\epsfbox{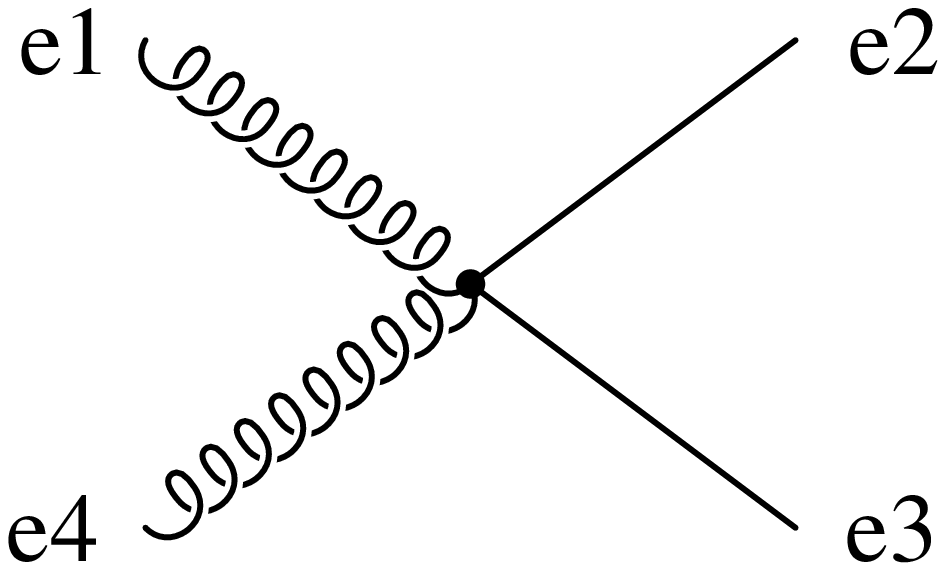}}
\end{picture}
}
&&&
i\GG{G^\mu_a G^\nu_b \epsilon \ygluibar_c} \nonumber\\*[1ex]&&{}={}&  i\sigma_{\mu\nu}gf_{cab}\\[2ex]
&&&
i\GG{G^\mu_a G^\nu_b \yglui_c \epsilonbar} \nonumber\\*[1ex]&&{}={}&  i\sigma_{\mu\nu}gf_{cab}\\[2ex]
\parbox[b]{\frwidth}
{
\begin{picture}(\frpicwidth,\frpicheight)
\psfrag{e1}{\parbox{7ex}{$\{\epsilonbar,\epsilon\}$\\[2.5ex]\mbox{\ }}}
\psfrag{e2}{$\sq^\dagger$}
\psfrag{e3}{$\sq$}
\psfrag{e4}{\parbox{7ex}{$\mbox{\ }\\[2.5ex]\{\yglui,\ygluibar\}$}}
\epsfxsize=6cm
\put(\frputleft,-120){\epsfbox{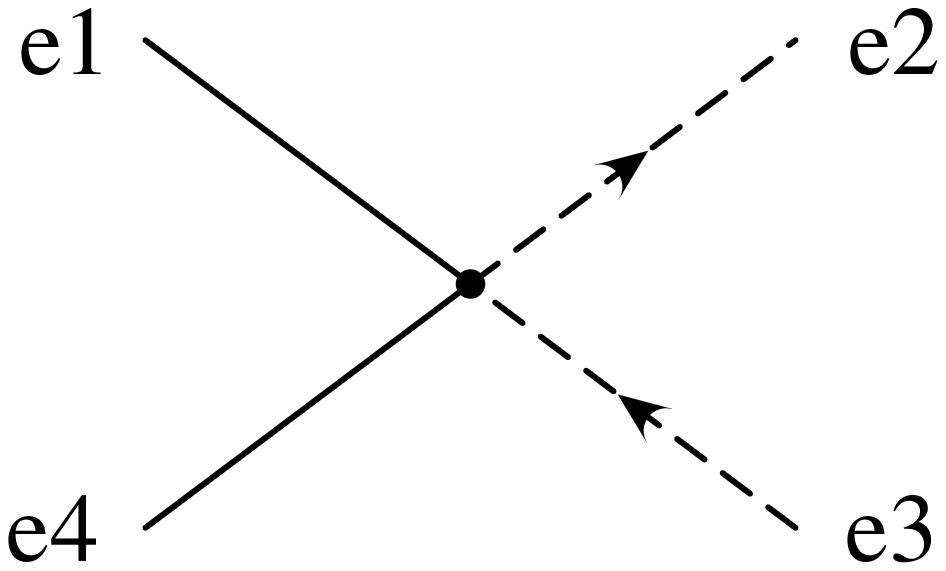}}
\end{picture}
}
&&&
i\GG{\yglui_a \epsilonbar \sq_{l,j} \sq^\dagger_{k,i}} \nonumber\\*[1ex]&&{}={}& 
 -ig(P_R-P_L)T^a_{ij}\times\nonumber\\* &&&(S_{kL}S^*_{lL} - S_{kR}S^*_{lR})\\[2ex]
&&&
i\GG{\epsilon \ygluibar_a  \sq_{l,j} \sq^\dagger_{k,i}} \nonumber\\*[1ex]&&{}={}& 
 +ig(P_R-P_L)T^a_{ij}\times\nonumber\\* &&&(S_{kL}S^*_{lL} - S_{kR}S^*_{lR})\\[2ex]
\parbox[b]{\frwidth}
{
\begin{picture}(\frpicwidth,\frpicheight)
\psfrag{e1}{$\yc$}
\psfrag{e2}{$c$}
\psfrag{e3}{$c$}
\epsfxsize=6cm
\put(\frputleft,-104){\epsfbox{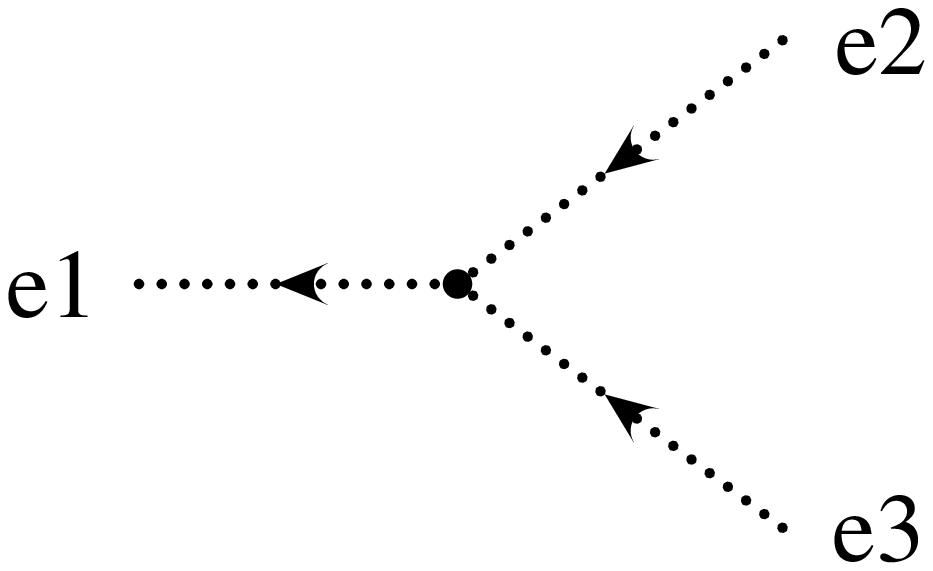}}
\end{picture}
}
&&&
i\GG{c_c c_b \yc_a} \nonumber\\*[1ex]&&{}={}&  igf_{abc}\\[2ex]
\parbox[b]{\frwidth}
{
\begin{picture}(\frpicwidth,\frpicheight)
\psfrag{e1}{$\epsilonbar$}
\psfrag{e2}{$\yc$}
\psfrag{e3}{$G$}
\psfrag{e4}{$\epsilon$}
\epsfxsize=6cm
\put(\frputleft,-104){\epsfbox{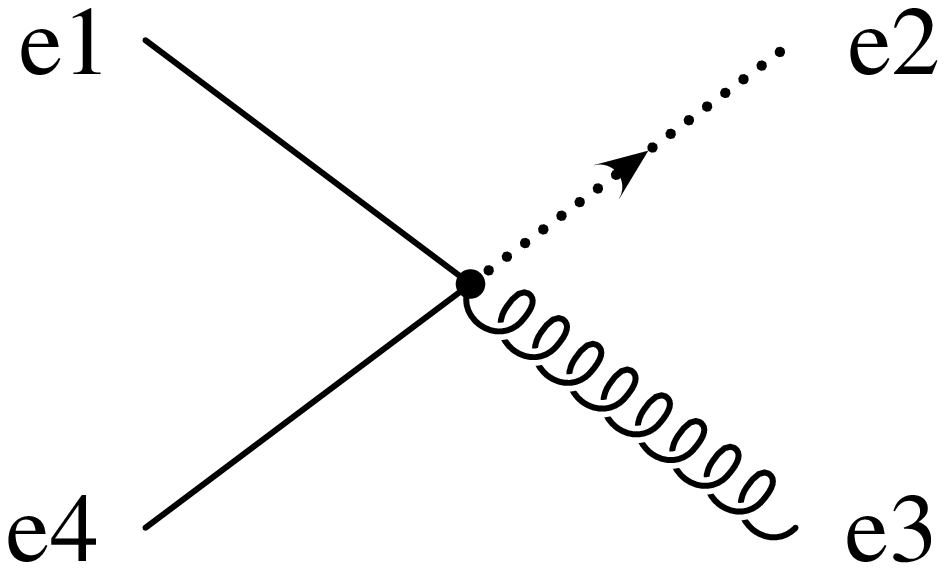}}
\end{picture}
}
&&&
i\GG{G^\mu_a \epsilon \epsilonbar \yc_b} \nonumber\\*[1ex]&&{}={}& 
 i 2 i \gamma_\mu \delta_{ab}\\[2ex]
\parbox[b]{\frwidth}
{
\begin{picture}(\frpicwidth,\frpicheight)
\psfrag{e1}{$\ysq$}
\psfrag{e2}{$\sq$}
\psfrag{e3}{$c$}
\epsfxsize=6cm
\put(\frputleft,-104){\epsfbox{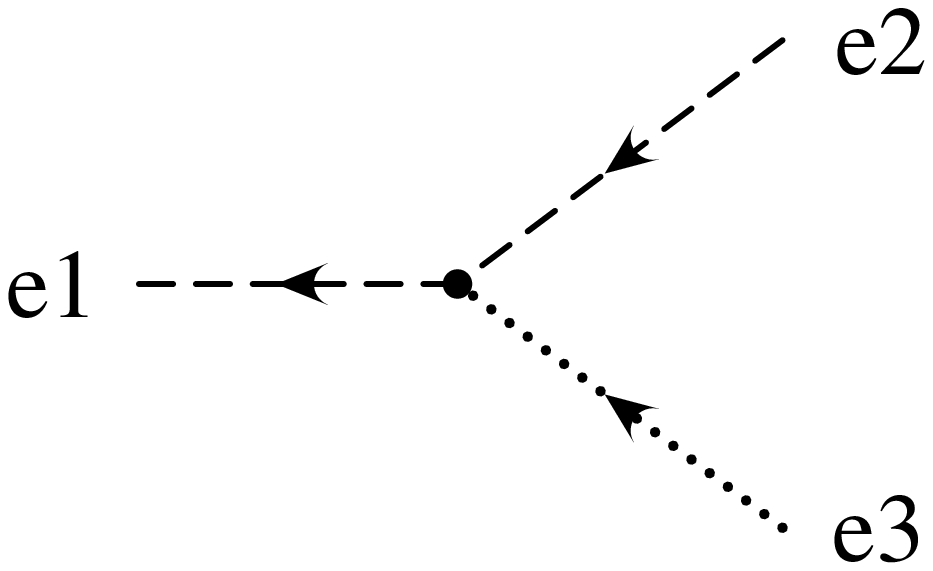}}
\end{picture}
}
&&&
i\GG{\sq_{k,j} c_a \ysq_{l,i}} \nonumber\\*[1ex]&&{}={}& g T^a_{ij}\delta_{kl}\\[2ex]
\parbox[b]{\frwidth}
{
\begin{picture}(\frpicwidth,\frpicheight)
\psfrag{e1}{$\ysq^\dagger$}
\psfrag{e2}{$\sq^\dagger$}
\psfrag{e3}{$c$}
\epsfxsize=6cm
\put(\frputleft,-104){\epsfbox{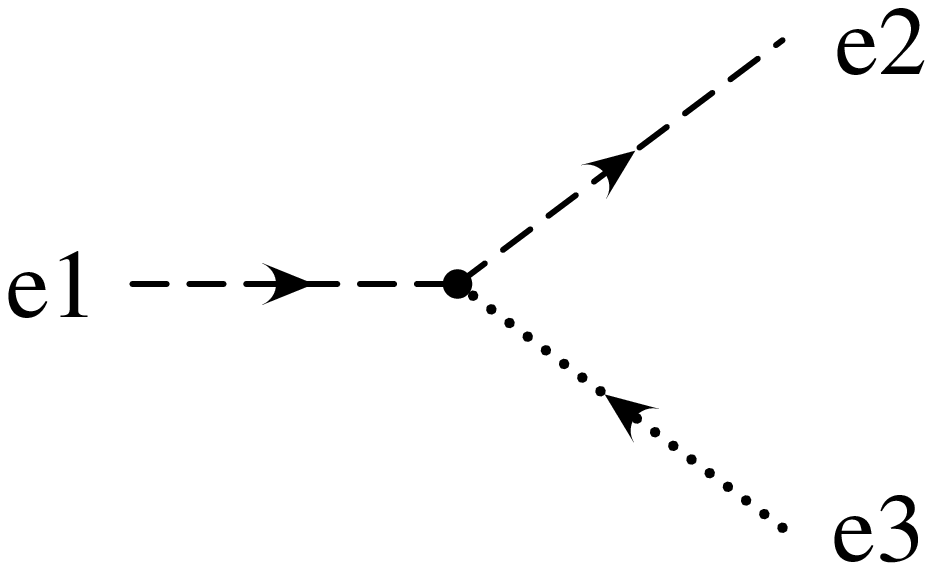}}
\end{picture}
}
&&&
i\GG{\ysq_{k,j}{}^\dagger c_a \sq_{l,i}^\dagger}
      \nonumber\\*[1ex]&&{}={}& -g T^a_{ij}\delta_{kl}\\[2ex]
\parbox[b]{\frwidth}
{
\begin{picture}(\frpicwidth,\frpicheight)
\psfrag{e1}{$\yq$}
\psfrag{e2}{$\qbar$}
\psfrag{e3}{$c$}
\epsfxsize=6cm
\put(\frputleft,-104){\epsfbox{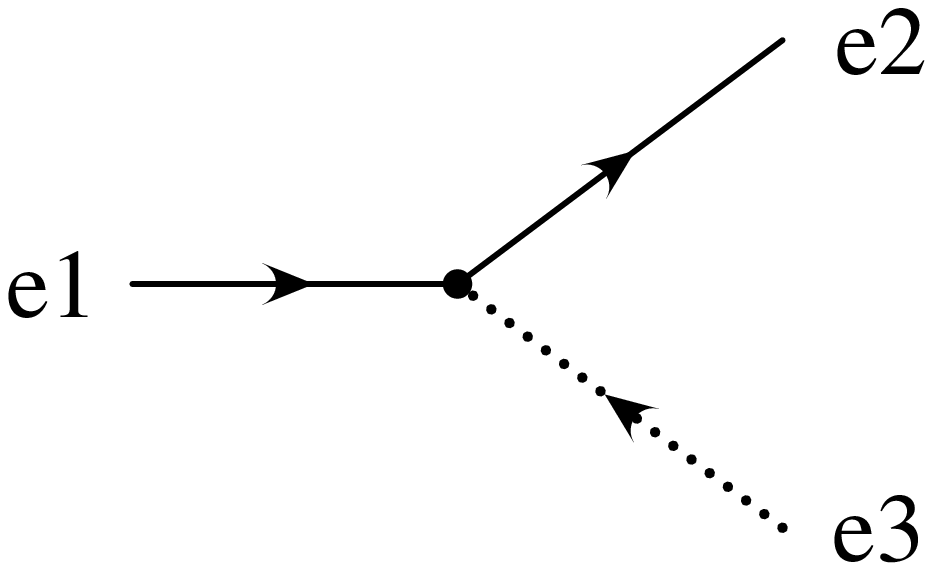}}
\end{picture}
}
&&&
i\GG{\yq_j c_a \qbar_i} \nonumber\\*[1ex]&&{}={}& -gT^a_{ij}\\[2ex]
\parbox[b]{\frwidth}
{
\begin{picture}(\frpicwidth,\frpicheight)
\psfrag{e1}{$\yqbar$}
\psfrag{e2}{$q$}
\psfrag{e3}{$c$}
\epsfxsize=6cm
\put(\frputleft,-104){\epsfbox{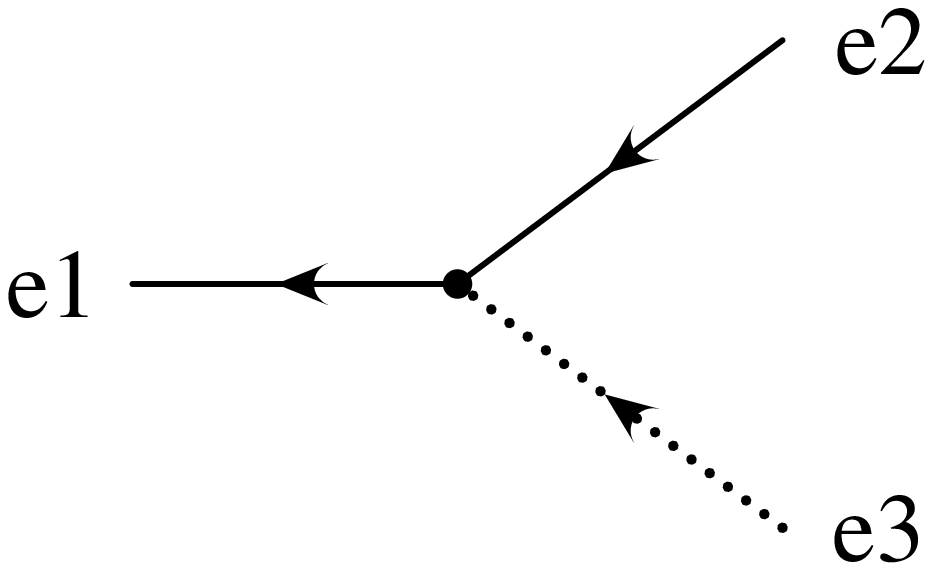}}
\end{picture}
}
&&&
i\GG{q_j c_a \yqbar_i} \nonumber\\*[1ex]&&{}={}& gT^a_{ij}\\[2ex]
\parbox[b]{\frwidth}
{
\begin{picture}(\frpicwidth,\frpicheight)
\psfrag{e1}{$\yqbar$}
\psfrag{e2}{$\sq$}
\psfrag{e3}{$G$}
\psfrag{e4}{$\epsilon$}
\epsfxsize=6cm
\put(\frputleft,-104){\epsfbox{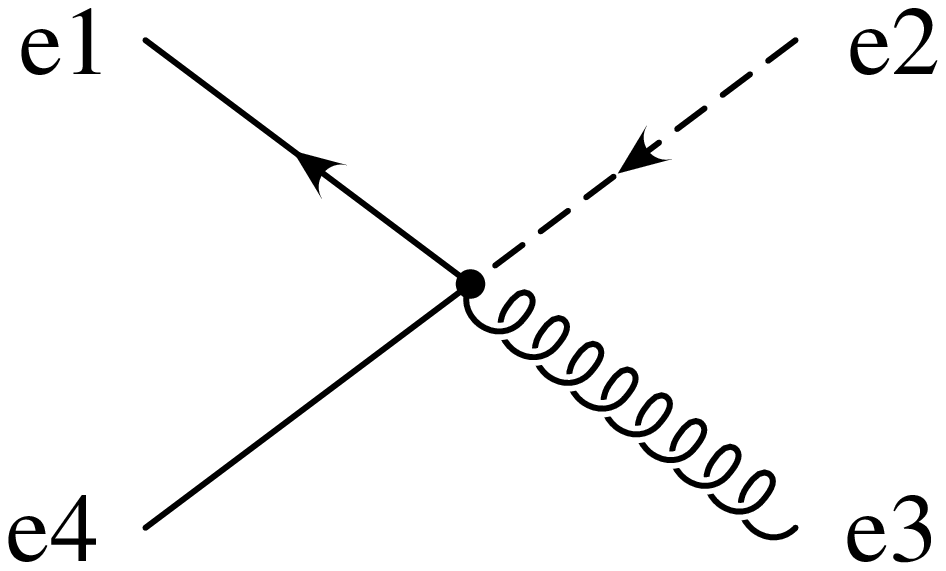}}
\end{picture}
}
&&&
i\GG{G^\mu_a \sq_{k,j} \epsilon \yqbar_i} \nonumber\\*[1ex]&&{}={}& 
 i\sqrt2 g \gamma_\mu T^a_{ij}\times\nonumber\\* &&&{}(-S^*_{kL}P_L+S^*_{kR}P_R)\\[2ex]
\parbox[b]{\frwidth}
{
\begin{picture}(\frpicwidth,\frpicheight)
\psfrag{e1}{$\yq$}
\psfrag{e2}{$\sq^\dagger$}
\psfrag{e3}{$G$}
\psfrag{e4}{$\epsilonbar$}
\epsfxsize=6cm
\put(\frputleft,-104){\epsfbox{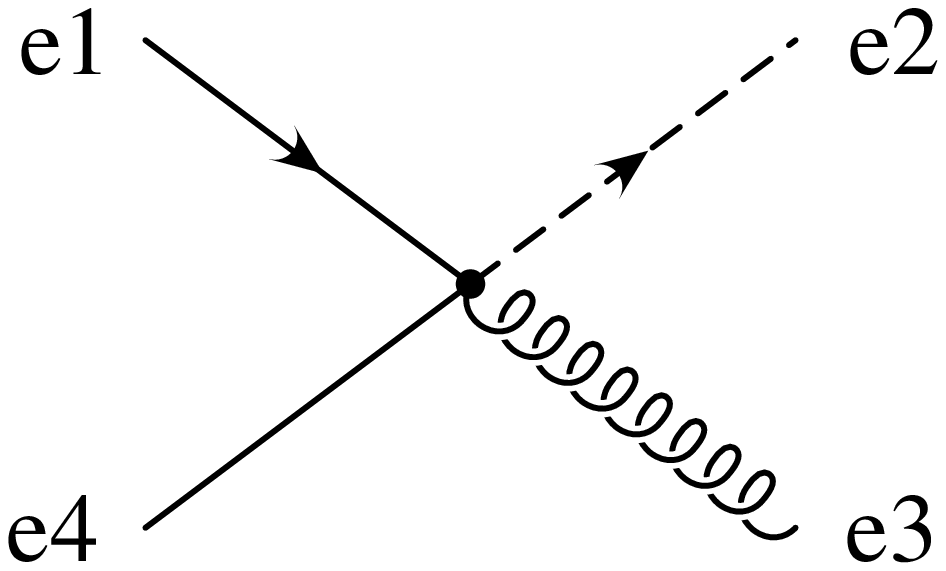}}
\end{picture}
}
&&&
i\GG{G^\mu_a \yq_j \sq^\dagger_{k,i} \epsilonbar} \nonumber\\*[1ex]&&{}={}& 
 i\sqrt2 g \gamma_\mu T^a_{ij}\times\nonumber\\* &&&{}(S_{kL}P_R - S_{kR}P_L)
\end{align*}

\section{One-loop results}
\label{Sec:Oneloop}
In the following we give a list of the one-loop Feynman diagrams and
results for the vertex functions corresponding to symmetry
transformations, i.e.\ vertex functions involving external $Y$ sources
and ghost fields. We use dimensional regularization with 
an anticommuting $\gamma_5$ or dimensional reduction and use the
variable $\dreg$ to distinguish both results. It takes the value
$\dreg=1$ in the case of dimensional regularization and $\dreg=0$ in
the case of dimensional reduction. \footnote{Here a word to the consistency
  of the schemes is in order. Both dimensional reduction and dimensional
  regularization in the way we use it are mathematically inconsistent
  and cannot be used at all orders. An inconsistent scheme can yield
  incorrect results if imaginary or non-local contributions turn out
  to be wrong, because this violates unitarity or causality. Here,
  however, it is easy to see that the difference of our schemes to a
  consistent one like the prescription of \cite{BM77} is a sum of
  local counterterms, and the results obtained using the scheme of
  \cite{BM77} with appropriate counterterms would coincide with
  ours. Therefore, our results are correct.}
\footnote{The Feynman gauge $\xi=1$ is used.} 
We specify the results in the limit of infinite momenta, which is
sufficient for our purposes. There all masses  and the subleading
momentum dependence can be neglected. In this subsection ${\cal O}(p^n)$
denotes a momentum dependence of the form $p^n\times\mbox{powers of
  }\log p$. The vertex functions
involving only physical fields can be calculated using standard
methods, so they are not displayed here.

The one-loop functions appearing are defined in
App.\ \ref{App1LInt} and have the arguments
\begin{eqnarray}
B_0 & = & B_0(p^2,0,0)\nonumber\ ,\\
C_0 & = & C_0(p^2,(p+k)^2,k^2,0,0,0)\nonumber\ ,\\
C_{1} & = & C_{1}(p^2,p^2,0,0,0,0)\ ,
\label{BCArgs}
\end{eqnarray}
where $p$ is the momentum argument of the corresponding vertex
function. Furthermore, as explained in sec.\ \ref{Sec:Parametrization}
we introduce non-symmetric counterterms $\delta_i$ to all vertex
functions except for the self energies and the $\sq\sq G_\mu$
interaction. 
These counterterms have to be chosen in such a way that the
Slavnov-Taylor identity is satisfied.


\subsection{Vertex functions involving $\ygluon^\mu, \yglui, \yc$}

\begin{figure}[htb]
\begin{center}
\begin{picture}(415,69)
\allpsfrag
\epsfxsize=8cm
\put(0,-80){\epsfbox{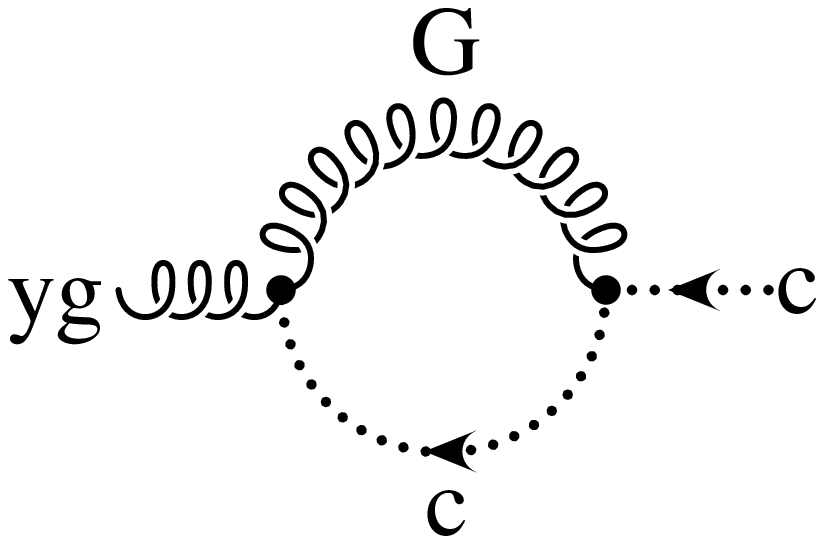}}
\end{picture}
\end{center}
\caption{The one-loop diagram contributing to the vertex function
  $\GG{\ygluon^\mu c}$.}
\label{FigcYg}
\end{figure}

\begin{figure*}[htb]
\begin{center}
\begin{picture}(415,69)
\allpsfrag
\epsfxsize=6cm
\put(-60,-60){\epsfbox{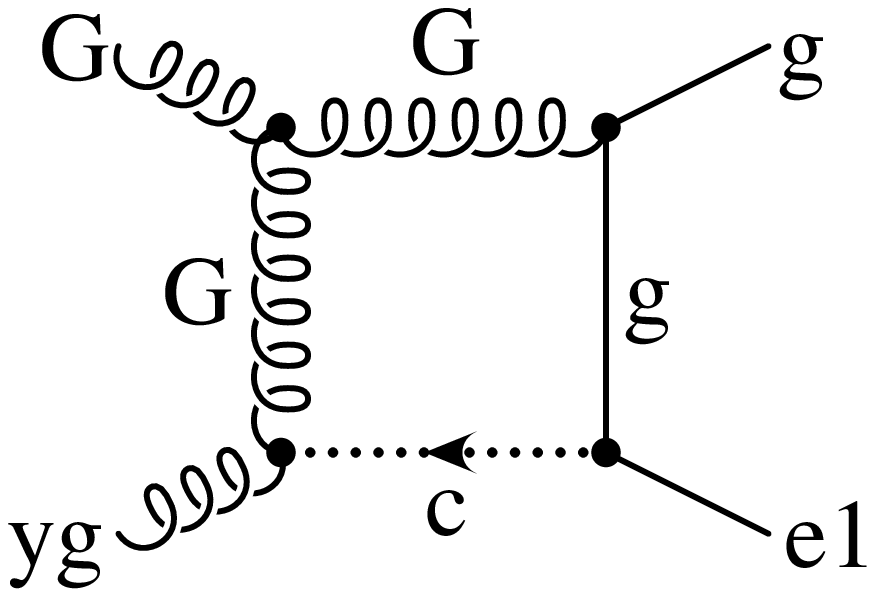}}
\epsfxsize=6cm
\put(90,-60){\epsfbox{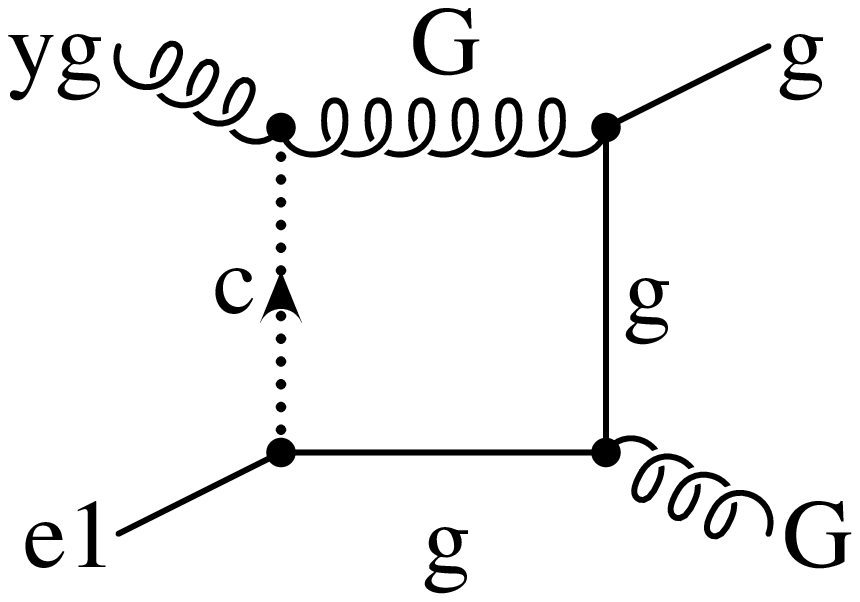}}
\epsfxsize=6cm
\put(240,-60){\epsfbox{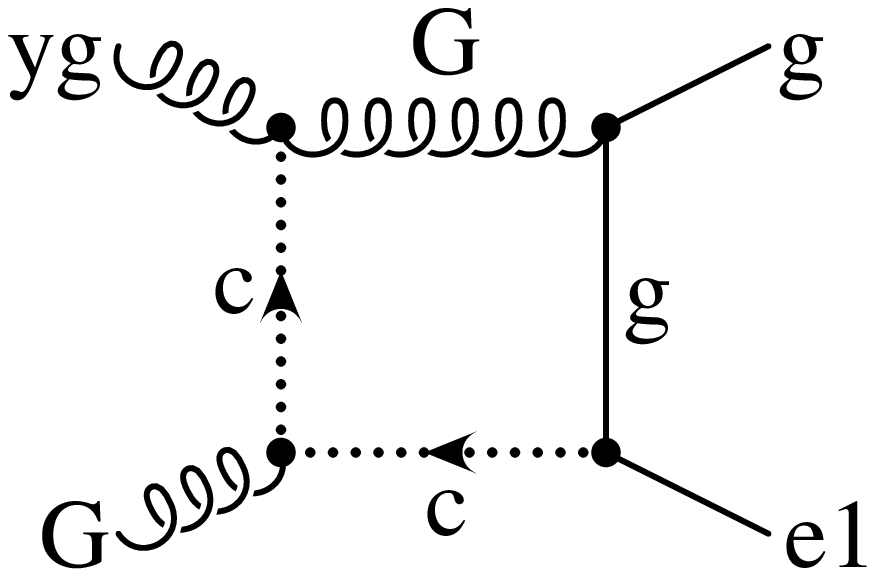}}
\end{picture}
\end{center}
\caption{The one-loop diagrams contributing to the vertex function
  $\GG{G^\mu\glui\epsilonbar\ygluon^\nu}$.}
\label{FigEpsYGgG}
\end{figure*}

\begin{figure}[htb]
\begin{center}
\begin{picture}(415,69)
\allpsfrag
\epsfxsize=6cm
\put(-40,-60){\epsfbox{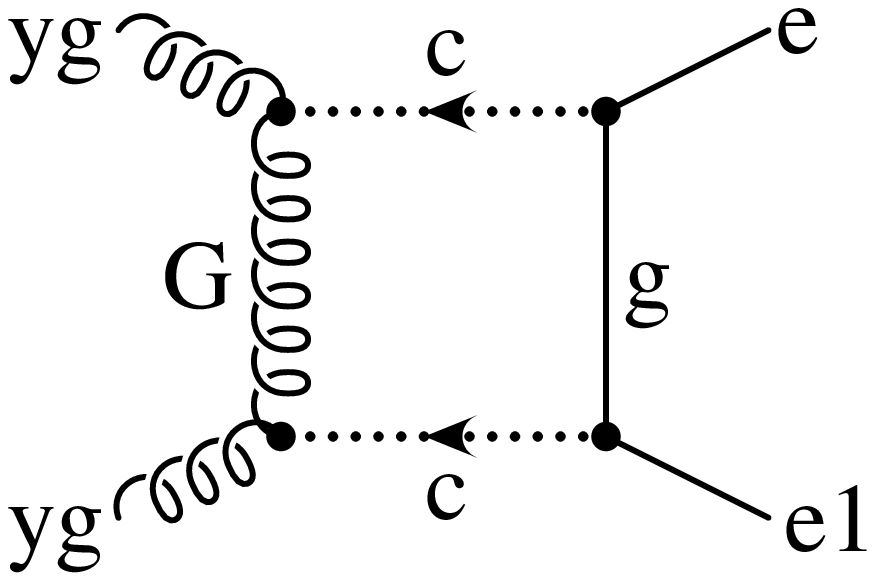}}
\epsfxsize=6cm
\put(100,-60){\epsfbox{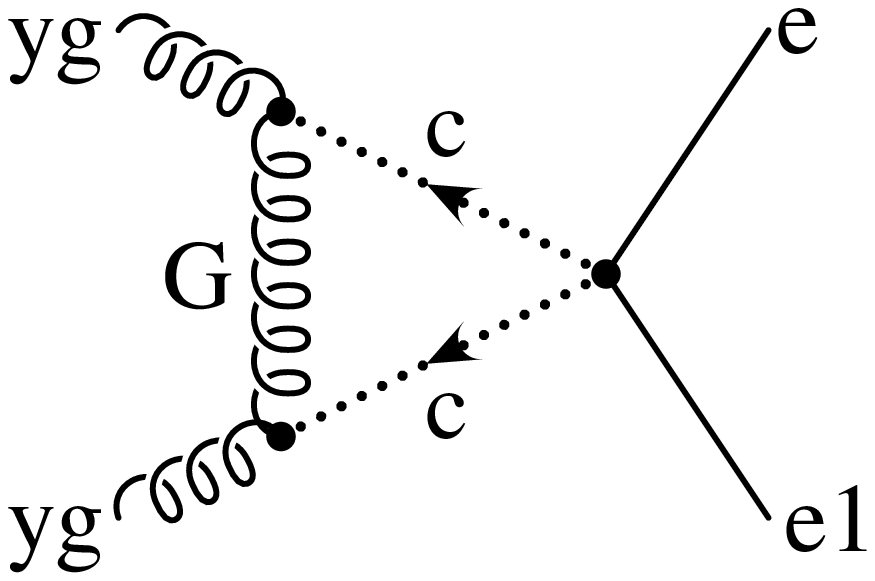}}
\end{picture}
\end{center}
\caption{The one-loop diagrams contributing to the vertex function
  $\GG{\ygluon^\mu\ygluon^\nu\epsilon\epsilonbar}$.}
\label{FigEpsYgYg}
\end{figure}

\begin{figure*}[htb]
\begin{center}
\begin{picture}(415,69)
\allpsfrag
\epsfxsize=6cm
\put(-40,-60){\epsfbox{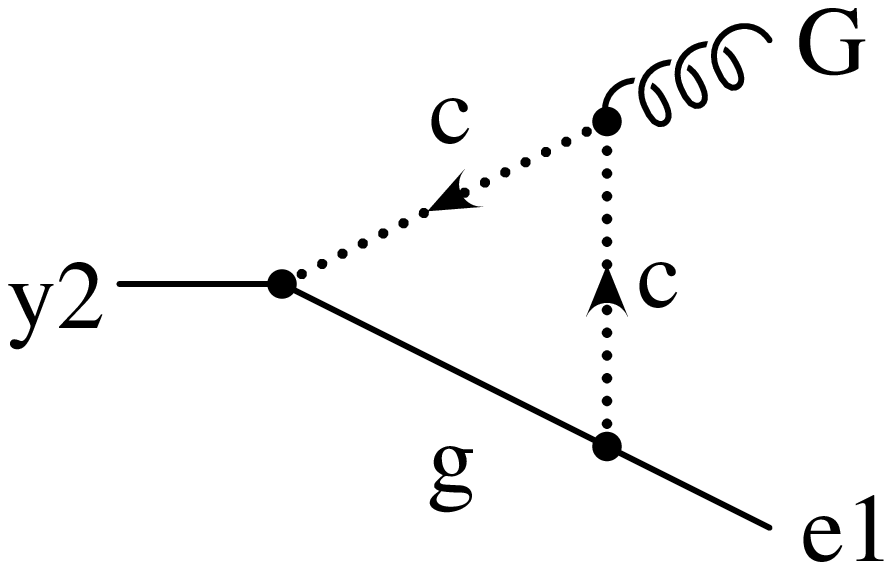}}
\epsfxsize=6cm
\put(70,-60){\epsfbox{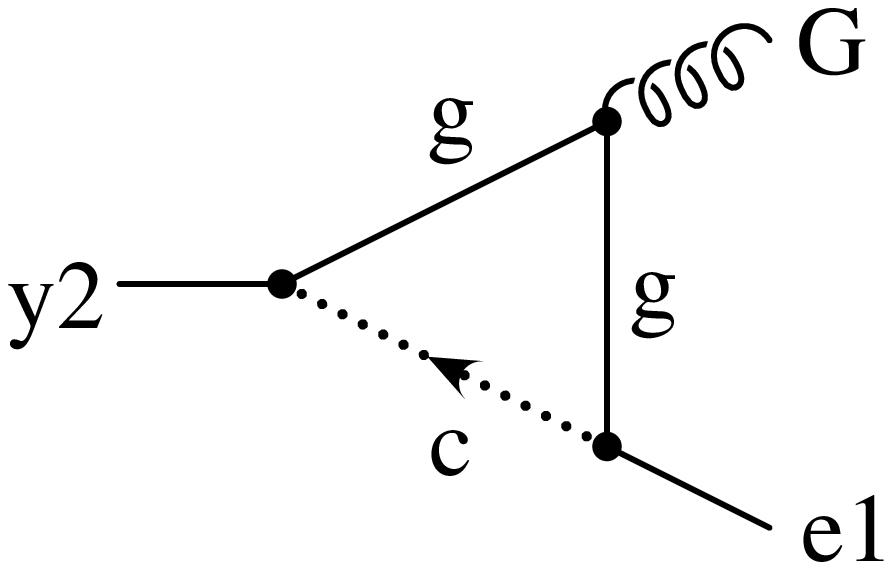}}
\epsfxsize=6cm
\put(180,-60){\epsfbox{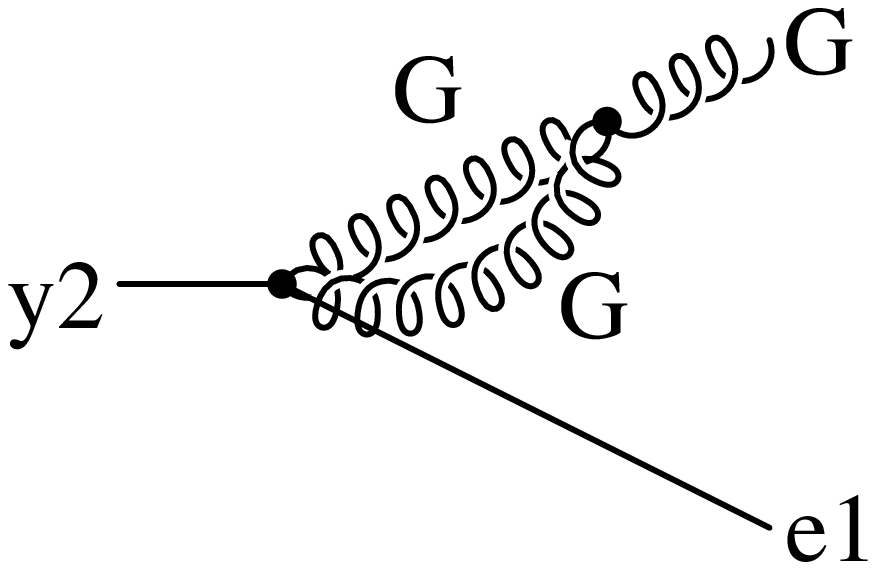}}
\epsfxsize=6cm
\put(290,-60){\epsfbox{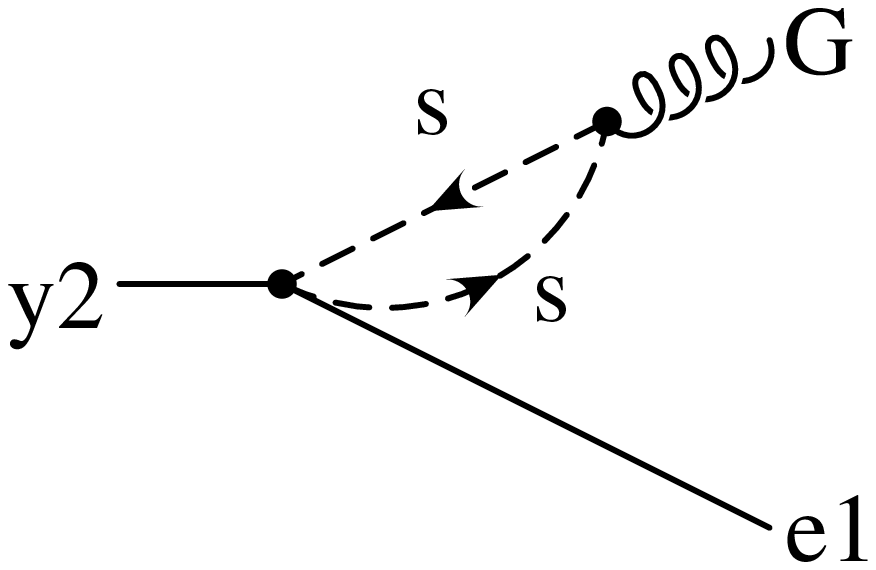}}
\end{picture}
\end{center}
\caption{The one-loop diagrams contributing to the vertex function
  $\GG{G^\nu \yglui\epsilonbar}$.}
\label{FigEpsYglui}
\end{figure*}

\begin{figure*}[htb]
\begin{center}
\begin{picture}(415,129)
\allpsfrag
\epsfxsize=6cm
\put(-40,10){\epsfbox{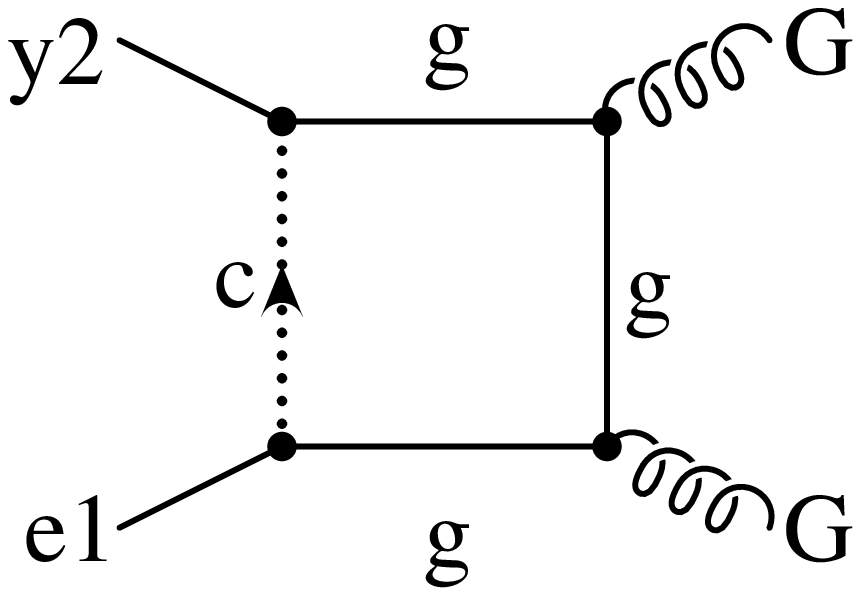}}
\epsfxsize=6cm
\put(70,10){\epsfbox{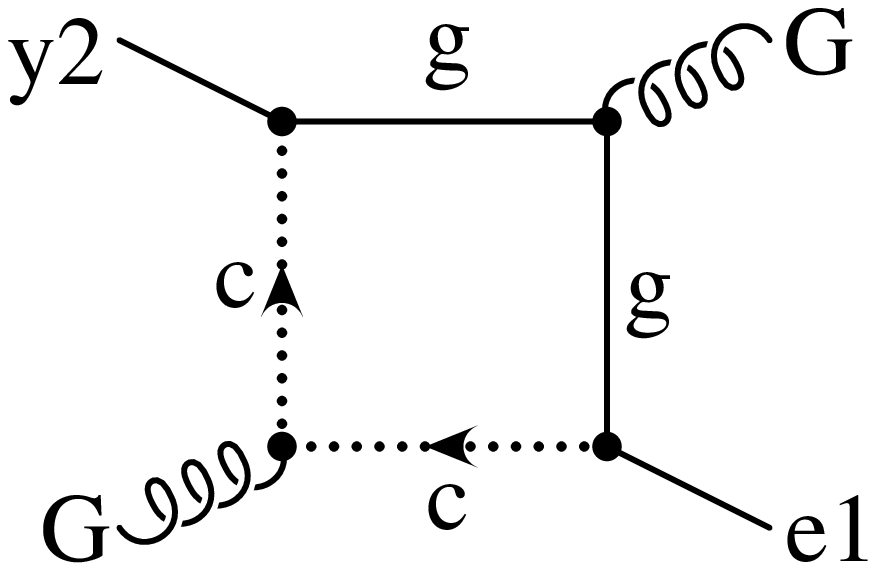}}
\epsfxsize=6cm
\put(180,10){\epsfbox{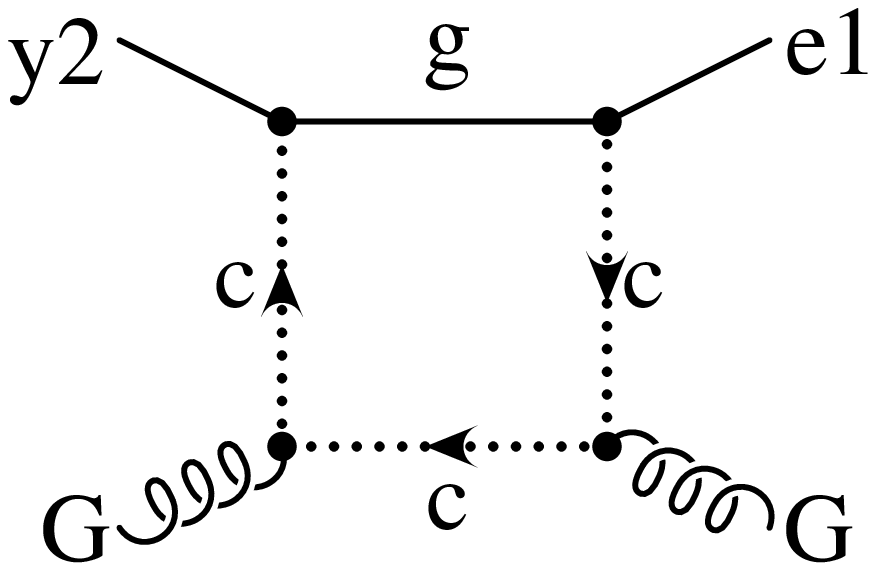}}
\epsfxsize=6cm
\put(290,10){\epsfbox{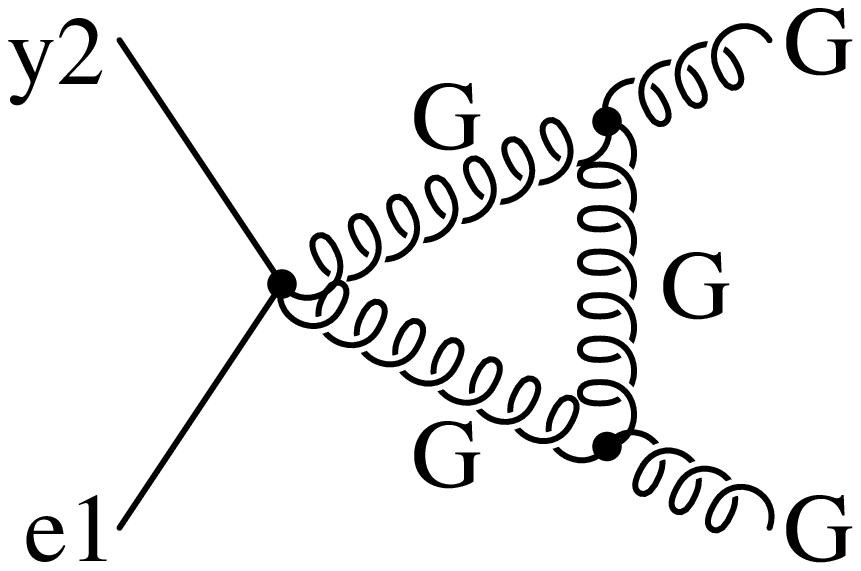}}
\epsfxsize=6cm
\put(-40,-60){\epsfbox{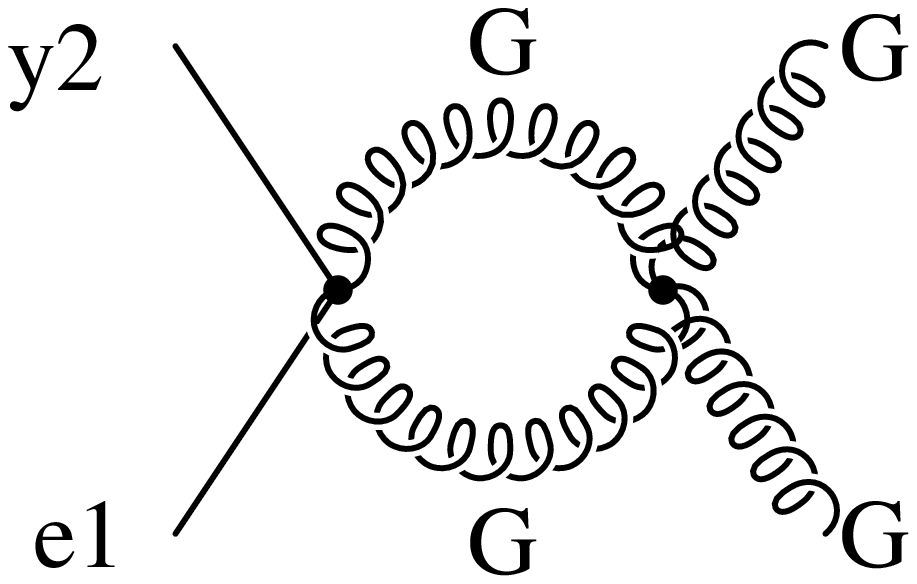}}
\epsfxsize=6cm
\put(70,-60){\epsfbox{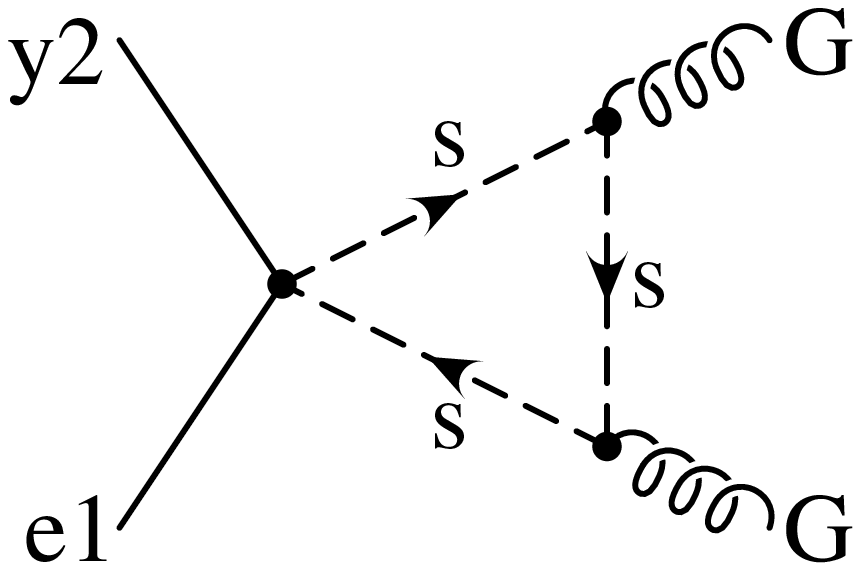}}
\epsfxsize=6cm
\put(180,-60){\epsfbox{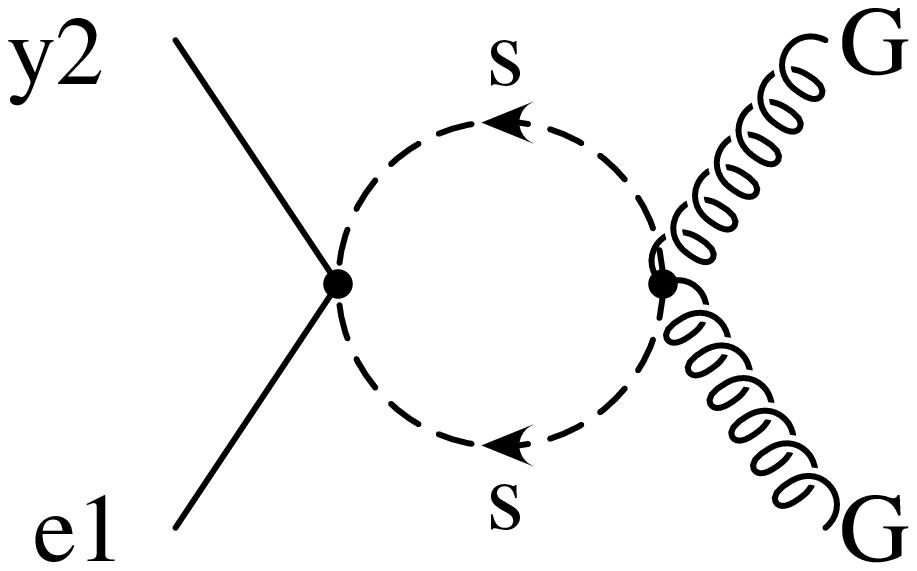}}
\end{picture}
\end{center}
\caption{The one-loop diagrams contributing to the vertex function
  $\GG{G^\mu G^\nu \yglui\epsilonbar}$.}
\label{FigEpsYgluiGG}
\end{figure*}

\begin{figure*}[htb]
\begin{center}
\begin{picture}(415,69)
\allpsfrag
\epsfxsize=6cm
\put(-60,-60){\epsfbox{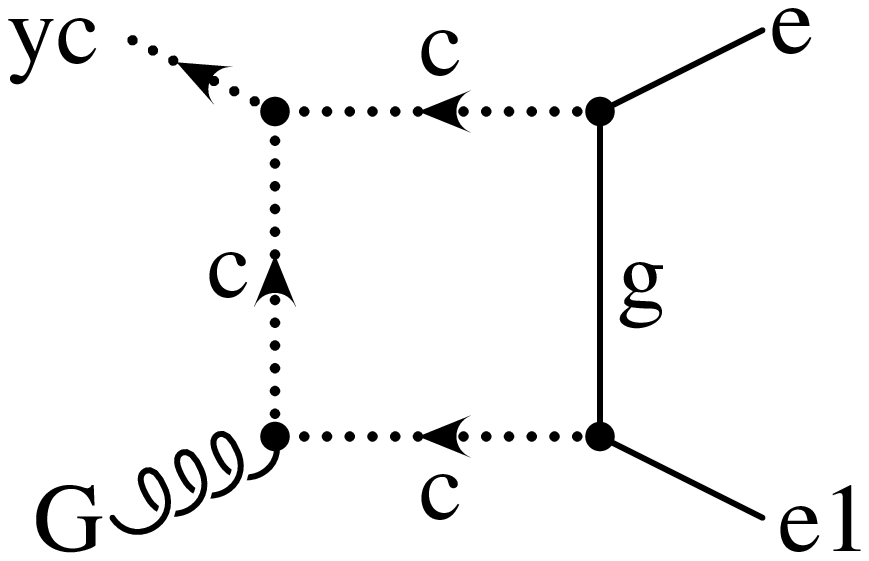}}
\epsfxsize=6cm
\put(90,-60){\epsfbox{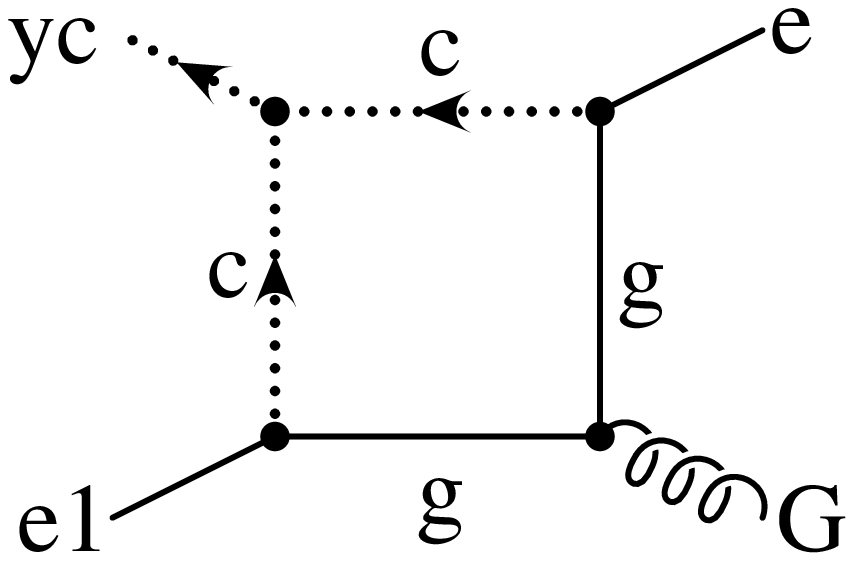}}
\epsfxsize=6cm
\put(240,-60){\epsfbox{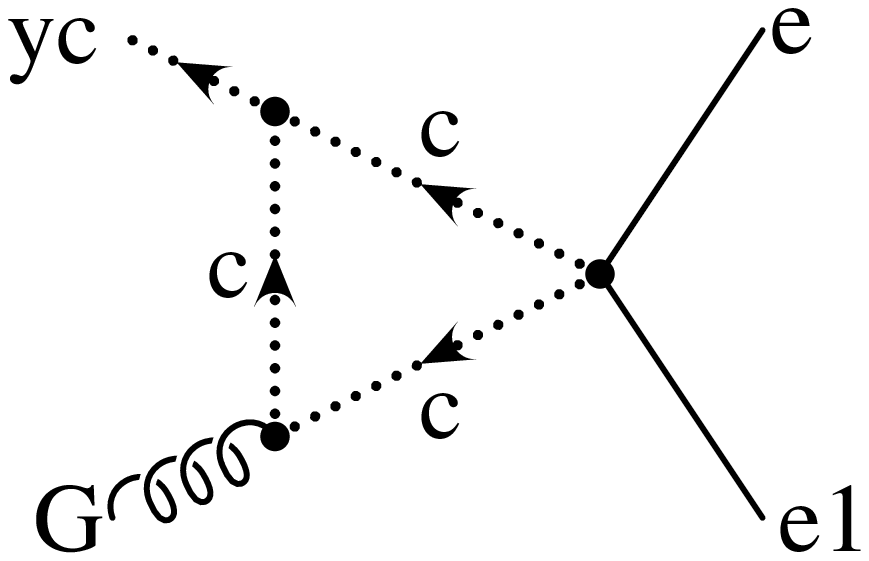}}
\end{picture}
\end{center}
\caption{The one-loop diagrams contributing to the vertex function
  $\GG{G^\nu\epsilon\epsilonbar\yc}$.}
\label{FigEpsYc}
\end{figure*}

\begin{eqnarray}
\lefteqn{
\GG{\ygluon{}^\mu_a c_c}(q,-q)  =  -iq_\mu\delta_{ac}\times}
\nonumber\\&&{}
\left(1 + \delta_{c\ygluon} - \frac{\alpha_s \CA}{4\pi}\frac12 B_0\right)\ ,\\
\lefteqn{
\GG{\ygluon{}^\mu_a\epsilon\gluibar_c}  =  -\gamma_\mu\delta_{ac}
\left(1 + \delta_{\ygluon\glui\epsilon} + \frac{\alpha_s \CA}{4\pi}
       B_0\right)\ ,}\\
\lefteqn{
\GG{\ygluon{}^\mu_a\glui_c\epsilonbar}  =  -\gamma_\mu\delta_{ac}
\left(1 + \delta_{\ygluon\glui\epsilon} + \frac{\alpha_s \CA}{4\pi}
         B_0\right) \ ,}\\
\lefteqn{
\GG{\epsilon\epsilonbar \ygluon^{\mu}_a \ygluon^{\nu}_c}  =  
{\cal O}(p^{-2})
\ ,}\\
\lefteqn{
\GG{G^\nu_b\yglui_c\epsilonbar}(-q,q)  = 
i\sigma_{\nu\mu}q^\mu\delta_{bc} \times}
\nonumber\\&&{}
\left(1 + \delta_{G\epsilon\yglui}- \frac{\alpha_s \CA}{4\pi}
       B_0\right)\ ,\\
\lefteqn{\GG{G^\nu_b\epsilon\ygluibar_c}(-q,q)  = 
i\sigma_{\nu\mu}q^\mu\delta_{bc}\times}
\nonumber\\&&{}
\left(1 +\delta_{G\epsilon\yglui}- \frac{\alpha_s \CA}{4\pi}
         B_0\right)\ ,\\
\lefteqn{\GG{G^\mu_a G^\nu_b \yglui_d \epsilonbar}(-q,0,q)  =  
-\frac12 i g f_{abd} \times}
\nonumber\\&&
\Biggl( 
       \left(4 \frac{\CA\alpha_s}{4\pi} q^2 B_0' + 2
             + 2\delta_{\yglui\epsilon GG}\right)
         g_{\mu\nu}
\nonumber\\&&{}
 - \left(3\frac{\CA\alpha_s}{4\pi} q^2 B_0' + 2
             + 2\delta_{\yglui\epsilon GG}\right) 
         \gamma_\mu  \gamma_\nu 
\nonumber\\&&{}
 - \frac{\CA\alpha_s}{4\pi} B_0' 
        \left(2 \gamma_\nu  \qsl q_\mu + 2 q_\nu q_\mu - 
              3 \gamma_\mu  \qsl q_\nu\right)
\Biggr)\ ,\\
\lefteqn{\GG{G^\nu_b \glui_c \ygluon^\rho_d \epsilonbar}(0,-q,q)  = 
 \frac{i}{2}\,\frac{\CA\alpha_s}{4\pi}g\,B_0'\,
  f_{cbd}\,\times}
\nonumber\\&&{}
\left( \qsl g_{\nu\rho} - \gamma_\nu \qsl \gamma_\rho - 
    3 \gamma_\rho q_\nu + \gamma_\nu q_\rho \right)\ ,\\
\lefteqn{
\GG{G^\nu_b\epsilon\epsilonbar \yc}  =  2i\gamma_\nu
\left(1 +\delta_{G\epsilon\epsilonbar\yc}+\frac{\alpha_s \CA}{4\pi}
           \frac12 (B_0)\right)\ .}
\end{eqnarray}



\subsection{Vertex functions involving $\ysq$, $\yq$}

\begin{figure*}[htb]
\begin{center}
\begin{picture}(415,69)
\allpsfrag
\epsfxsize=6cm
\put(-60,-60){\epsfbox{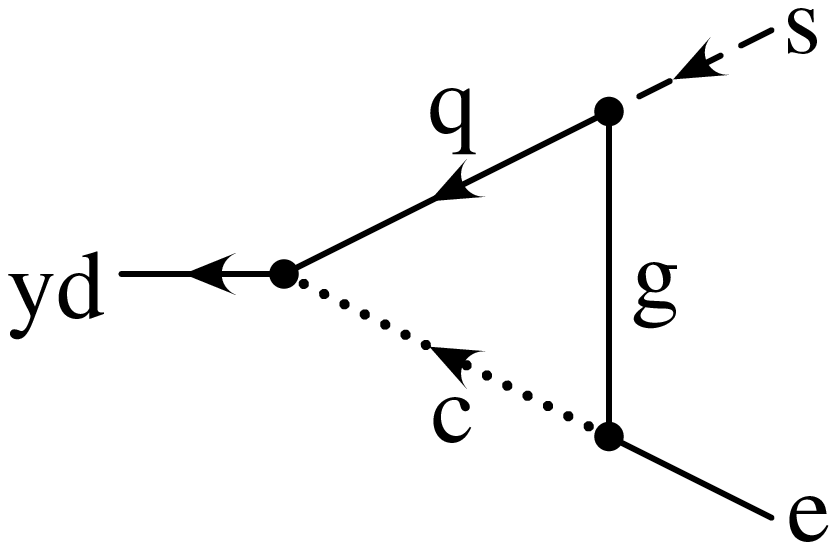}}
\epsfxsize=6cm
\put(90,-60){\epsfbox{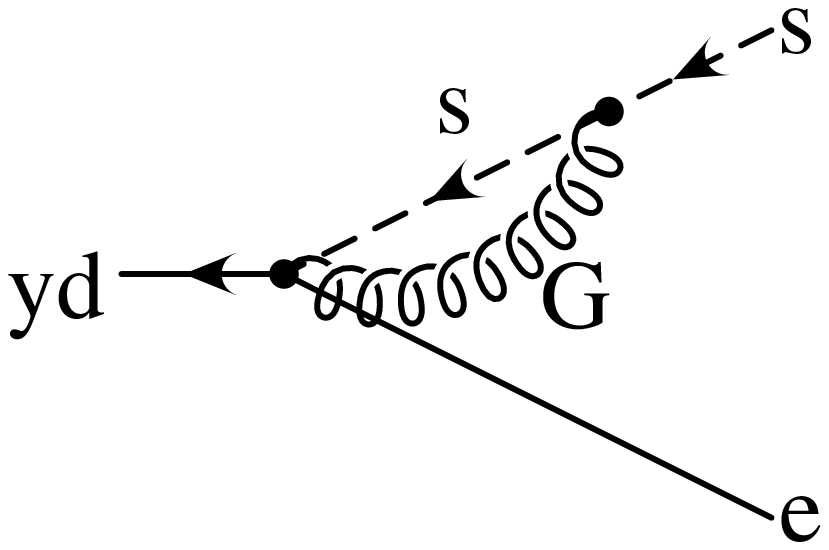}}
\epsfxsize=6cm
\put(240,-60){\epsfbox{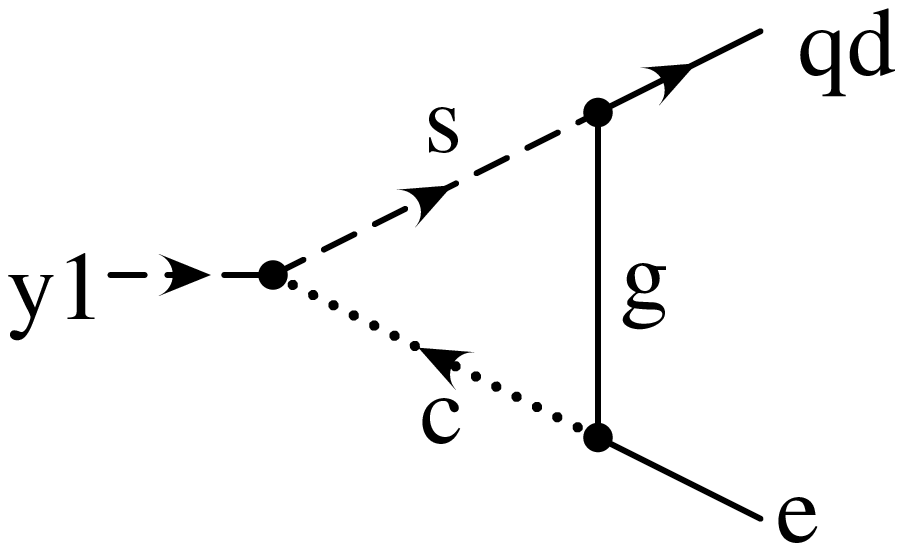}}
\end{picture}
\end{center}
\caption{The one-loop diagrams contributing to the vertex functions
  $\GG{\sq\epsilon\yqbar}$ and
  $\GG{\epsilon\qbar\ysq^\dagger}$.}  
\label{FigEpsYq}
\end{figure*}

\begin{figure*}[htb]
\begin{center}
\begin{picture}(415,69)
\allpsfrag
\epsfxsize=6cm
\put(-40,-60){\epsfbox{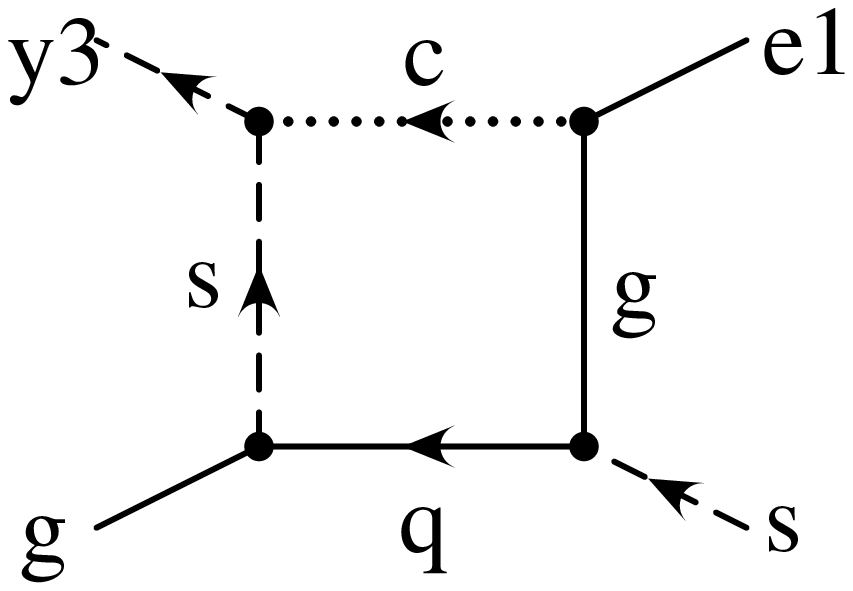}}
\epsfxsize=6cm
\put(70,-60){\epsfbox{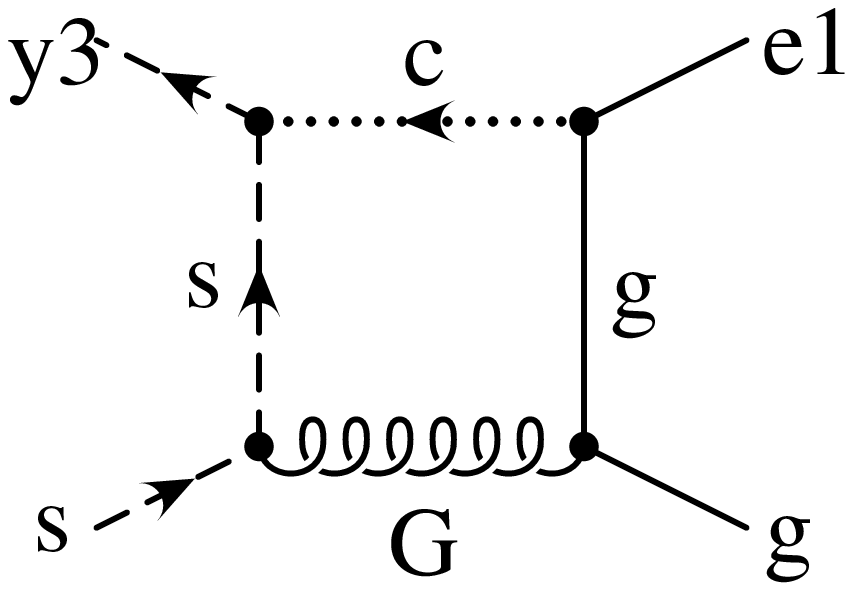}}
\epsfxsize=6cm
\put(180,-60){\epsfbox{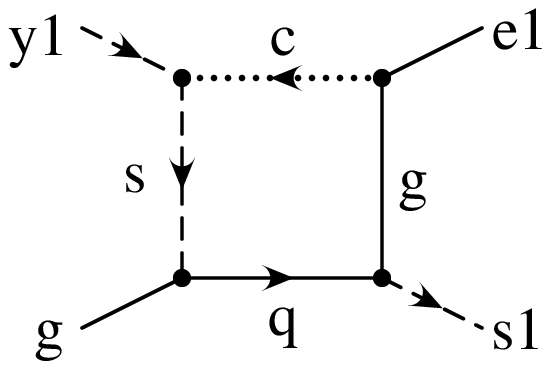}}
\epsfxsize=6cm
\put(290,-60){\epsfbox{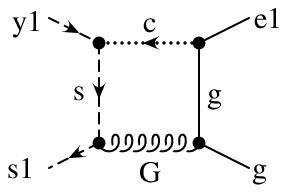}}
\end{picture}
\end{center}
\caption{The one-loop diagrams contributing to the vertex functions
  $\GG{\sq\glui\epsilonbar\ysq}$ and
  $\GG{\sq^\dagger\glui\epsilonbar\ysq^\dagger}$.}  
\label{FigEpsYsq}
\label{FirstDiagram}
\end{figure*}

\begin{figure}[htb]
\begin{center}
\begin{picture}(415,69)
\allpsfrag
\epsfxsize=6cm
\put(-40,-60){\epsfbox{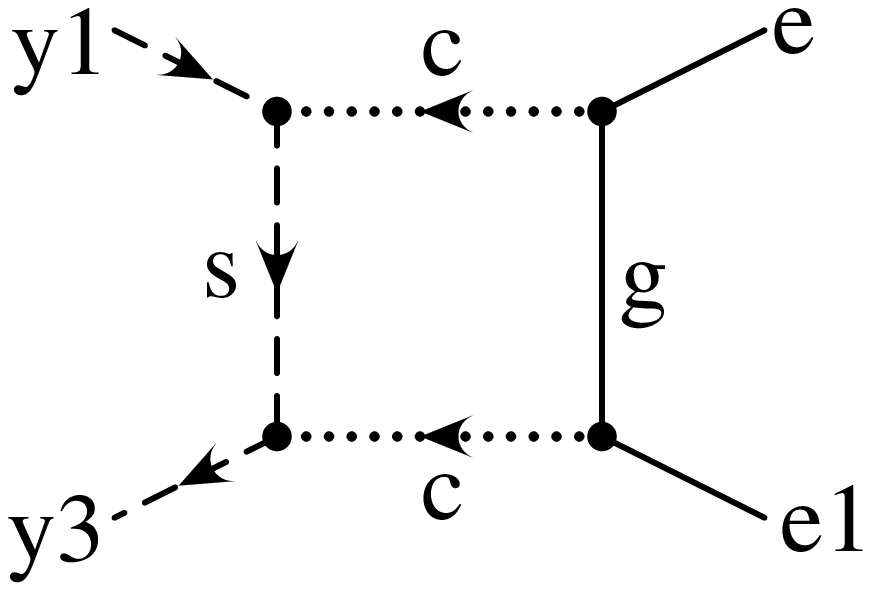}}
\epsfxsize=6cm
\put(100,-60){\epsfbox{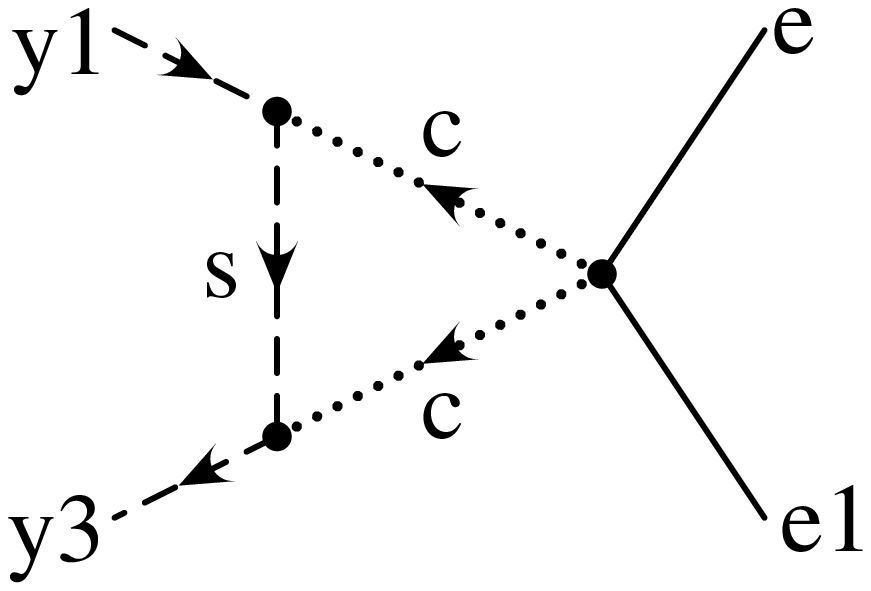}}
\end{picture}
\end{center}
\caption{The one-loop diagrams contributing to the vertex function
  $\GG{\ysq\ysq^\dagger\epsilon\epsilonbar}$.}
\label{FigEpsYsqYsq}
\label{LastDiagram}
\end{figure}

\begin{eqnarray}
\lefteqn{
\GG{\ysq^\dagger_{k,j}\epsilon \qbar_i}  =  \sqrt2
 (S^*_{kL}P_R - S^*_{kR}P_L)\delta_{ij}\times}
\nonumber\\&&{}
 \left(1+\delta_{\ysq\epsilon q}+\frac{\alpha_s \CF}{4\pi}B_0\right)\ ,\\
\lefteqn{
\GG{q_i\epsilonbar\ysq_{k,j}}  =  \sqrt2
 (S_{kL}P_L - S_{kR}P_R)\delta_{ij}\times}
\nonumber\\&&{}
 \left(1+\delta_{\ysq\epsilon q}+\frac{\alpha_s \CF}{4\pi}B_0\right)\ ,\\
\lefteqn{
\GG{\sq_{k,j}\epsilon \yqbar_i}(q,-q)  =  \sqrt2
 \qsl\, (S^*_{kL}P_R - S^*_{kR}P_L)\delta_{ij}\times}
\nonumber\\&&{}
 \left(1+\delta_{\yq\sq\epsilon}-\frac{\alpha_s \CF}{4\pi}(B_0)\right)\ ,\\
\lefteqn{
\GG{\sq_{k,j}^\dagger \yq_i\epsilonbar}(-q,q)  =  \sqrt2
 \qsl (S_{kL}P_R - S_{kR}P_L)\delta_{ij}\times}
\nonumber\\&&{}
 \left(1+\delta_{\yq\sq\epsilon}-\frac{\alpha_s \CF}{4\pi}(B_0)\right)\ ,\\
\lefteqn{
\GG{\sq_L\glui_a\epsilonbar\ysq_L}(p,k,-p-k)  = 
g\,\psl\, T^a \frac{\alpha_s}{4\pi}\times}
\nonumber\\&&{}
 \Bigl(\frac{\CA }{2}(2C_0+(1-2P_L)C_{1})
\nonumber\\&&{}
 + \CF  (2 P_L C_{1})\Bigr)
+{\cal O}(k^{-1})
\ ,\\
\lefteqn{
\GG{\sq_L^\dagger\glui_a\epsilonbar\ysq_L^{\dagger}}(-p,k,p-k)  = 
-g\,\psl\, T^a \frac{\alpha_s}{4\pi}\times}
\nonumber\\&&{}
 \Bigl(\frac{\CA }{2}(2C_0+(1-2P_R)C_{1})
\nonumber\\&&{}
 + \CF  (2 P_R C_{1})\Bigr)
+{\cal O}(k^{-1})
\ ,\\
\lefteqn{
\GG{\ysq_j\epsilon\epsilonbar\ysq^\dagger_i}(q,-q)  = 
 {\cal O}(q^{-2})
\ .}
\end{eqnarray}

\subsection{Identities involving $\omega^\mu$}
Owing to the non-renormalization of the terms involving $\omega^\mu$
(see sec.\ \ref{Sec:Definition}) we have
\begin{eqnarray}
\GG{\sq_j\ysq_i\omega^\mu}(q,-q) & = & \delta_{ij}q_\mu\ ,\\
\GG{G^\rho_a\ygluon^\sigma_b \omega^\mu}(q,-q) & = &
g_{\rho\sigma}\delta_{ab} q_\mu\ ,\\
\frac{\delta^2 s\omega^\mu}{\delta\epsilon\delta\epsilonbar}
 & = & 2\gamma^\mu\ .
\end{eqnarray}

\section{One-loop functions}
\label{App1LInt}
We use the following one-loop two- and three-point
functions \cite{PaVe79}:
\begin{eqnarray}
B_{0} & = & \int
        \frac{1}{[k^2-m_0^2][(k+p_1)^2-m_1^2]}
\ ,\\
C_{\{0,\mu\}} & = & \int
        \frac{\{1,k_\mu\}}
             {\scriptstyle[k^2-m_0^2][(k+p_1)^2-m_1^2][(k+p_2)^2-m_2^2]}
\ ,
\end{eqnarray}
with
\begin{eqnarray}
\int & \to & \mu^{4-D}\frac{16\pi^2}{i}\int \frac{d^Dk}{(2\pi)^D}
\end{eqnarray}
and the tensor decomposition
\begin{eqnarray}
C_\mu & = & {p_1}_\mu C_{1} + {p_2}_\mu C_{2}
\ ,\\
B_{0} & = & B_{0}(p_1^2,m_0^2,m_1^2)
\ ,\\
C_{ij} & = & C_{ij}(p_1^2,(p_2-p_1)^2,p_2^2,m_0^2,m_1^2,m_2^2)
\end{eqnarray}
in the conventions of \cite{LoopTools,Denner93}.

\section{Useful formulas}

\subsection{$SU(3)$}
\label{Sec:SU3}

\begin{align}
[T^a,T^b] & =  if_{abc}T^c\ ,\\
f_{abc}f_{dbc} & =  \CA  \delta_{ad}\ ,&\CA  & =  3\ ,\\
{\rm Tr}(T^aT^b) & =  \TF  \delta_{ab}\ ,&\TF & =  \frac12\ ,\\
(T^aT^a)_{ij} & =  \CF  \delta_{ij}\ ,&\CF  & =  \frac43
\ .
\end{align}

\subsection{Spinor identities}

In $\L_{\rm ext}$ and the Slavnov-Taylor operator several useful
replacements are possible. The signs are due to the bosonic statistics
of the external spinors $\yglui$, $\yq$: 
\begin{eqnarray}
\ygluibar s\glui & = & - (s\gluibar)\yglui\ ,\\
\dg{\glui_a}\dg{\ygluibar_a} & = & -\dg{\yglui_a}\dg{\gluibar_a}\ ,\\
\dg{q}\dg{\yqbar} & = & -\dg{\yq {}^C}\dg{\qbar^C}\ ,\\
\dg{\yq}\dg{\qbar} & = & -\dg{q^C}\dg{\yqbar{}^C}\ .
\end{eqnarray}


\end{appendix}

\begin{flushleft}

\end{flushleft}
\end{document}